\documentclass[12pt,letterpaper]{article}
\pdfoutput=1
\usepackage{graphicx,array}
\usepackage{color}
\usepackage{latexsym}
\usepackage{amsthm}
\usepackage{amsmath}
\usepackage{amssymb}
\usepackage{hyperref}
\usepackage{bbold}
\usepackage{subfig}
\usepackage{cancel}

\numberwithin{equation}{section}

\def\simgt{\mathrel{\lower2.5pt\vbox{\lineskip=0pt\baselineskip=0pt
           \hbox{$>$}\hbox{$\sim$}}}}
\def\simlt{\mathrel{\lower2.5pt\vbox{\lineskip=0pt\baselineskip=0pt
           \hbox{$<$}\hbox{$\sim$}}}}

\newcommand{\be}{\begin{equation}}
\newcommand{\ee}{\end{equation}}
\newcommand{\bea}{\begin{eqnarray}}
\newcommand{\eea}{\end{eqnarray}}
\newcommand{\Eq}[1]{Eq.~(\ref{#1})}
\newcommand{\Eqs}[2]{Eqs.~(\ref{#1}) and (\ref{#2})}

\newcommand{\MeV}{\textrm{ MeV}}
\newcommand{\GeV}{\textrm{ GeV}}
\newcommand{\TeV}{\textrm{ TeV}}
\newcommand{\gsim}{\lower.7ex\hbox{$\;\stackrel{\textstyle>}{\sim}\;$}}
\newcommand{\lsim}{\lower.7ex\hbox{$\;\stackrel{\textstyle<}{\sim}\;$}}

\newcommand{\bPhi}{{\boldsymbol \Phi}}

\newcommand{\bA}{{\boldsymbol A}}
\newcommand{\bK}{{\boldsymbol K}}
\newcommand{\bW}{{\boldsymbol W}}
\newcommand{\bX}{{\boldsymbol X}}
\newcommand{\bS}{{\boldsymbol S}}
\newcommand{\bHu}{{\boldsymbol H_u}}
\newcommand{\bHd}{{\boldsymbol H_d}}

\newcommand{\bH}{{\boldsymbol H}}
\newcommand{\bZ}{{\boldsymbol Z}}

\newcommand{\beff}{{\boldsymbol f}}
\newcommand{\bSigma}{{\boldsymbol \Sigma}}
\newcommand{\bsigma}{{\boldsymbol \sigma}}

\hypersetup{colorlinks,citecolor= blue,linkcolor= black}

\begin{document}

SCIPP 16/05 \\

\hfill

\vspace{1.5cm}

\begin{center}
{\LARGE\bf
SaxiGUTs and their Predictions
}
\\ \vspace*{0.5cm}

\bigskip\vspace{1cm}{
{\large \mbox{Raymond T. Co$^{1,2}$, Francesco D'Eramo$^{3,4}$ and Lawrence J. Hall$^{1,2}$} }
} \\[7mm]
{\it 
$^1$Berkeley Center for Theoretical Physics, Department of Physics, \\ University of California, Berkeley, CA 94720, USA \\
$^2$Theoretical Physics Group, Lawrence Berkeley National Laboratory, Berkeley, CA 94720, USA \\
$^3$Department of Physics, University of California Santa Cruz, Santa Cruz, CA 95064, USA \\
$^4$Santa Cruz Institute for Particle Physics, Santa Cruz, CA 95064, USA} 
\end{center}

\bigskip
\centerline{\large\bf Abstract}

\vspace{0.3cm}

\begin{quote} \small
We introduce a class of unified supersymmetric axion theories with unified and PQ symmetries broken by the same set of fields at a scale $\sim 2\times 10^{16}$ GeV.  A typical domain wall number of order 30 leads to an axion decay constant $f_a$ of order $10^{15}$ GeV.  Inflation generates a large saxion condensate giving a reheat temperature $T_R$ below the QCD scale for supersymmetry breaking of order $1-10$ TeV.  Axion field oscillations commence in the saxion matter-dominated era near the QCD scale, and recent lattice computations of the temperature dependence of the axion mass in this era allow a controlled calculation of the axion dark matter abundance.  A successful prediction of this abundance results for an initial axion misalignment angle of order unity, $\theta_i \sim 1$.  A highly correlated set of predictions is discussed for $f_a$, $T_R$, the supersymmetric Higgs mass parameter $\mu$, the amount of dark radiation $\Delta N_{eff}$, the proton decay rate $\Gamma(p \rightarrow e^+ \pi^0)$, isocurvature density perturbations and the $B$-mode of the cosmic microwave background.  The last two are particularly interesting when the energy scale of inflation is also of order $10^{16}$ GeV.

\end{quote}

\setcounter{tocdepth}{1}

\newpage
\tableofcontents

\section{Introduction}
\label{sec:intro}

An elegant solution to the strong CP problem was proposed in 1977  by promoting the strong CP parameter $\bar{\theta}$ to a field \cite{Peccei:1977hh,Weinberg:1977ma,Wilczek:1977pj}.   This field, the axion $a(x)$, is the pseudo-Goldstone boson produced by spontaneously breaking a global U(1) symmetry at scale $f_a$. The axion acquires a mass because this PQ symmetry is explicitly broken by a color anomaly, leading to an interaction in the low energy effective theory of 
\be
{\cal L}_{aG\tilde{G}} =\frac{g_3^2}{32\pi^2} \, \frac{a}{f_a} \, G^{\mu\nu} \tilde{G}_{\mu\nu}
\label{eq:f}
\ee
where $G$ is the QCD field strength.   By 1981, constraints from particle physics and astrophysics already required this scale to be very much larger than the weak scale, $f _a> 10^9$ GeV \cite{Dicus:1979ch}.  At that time it appeared natural to identify $f_a$ as the scale of grand unification, either non-supersymmetric \cite{Wise:1981ry} or supersymmetric \cite{Nilles:1981py}, with the same field breaking both the unified gauge and PQ symmetries. 

However, in 1983 axion production in the early Universe by the misalignment mechanism was discovered, suggesting a limit on $f_a$ far below the unified scale \cite{Preskill:1982cy, Abbott:1982af, Dine:1982ah}. Requiring that axions produced from oscillations of the misaligned condensate do not give more dark matter than observed today yields
\be 
f_a \theta_i^{1.7} \simlt 10^{12} \GeV \ ,
\label{eq:f83}
\ee
effectively decoupling PQ breaking from physics at the unified scale. For the Pre-Inflation cosmology, where PQ breaking occurs before inflation, $\theta_i$ is the initial axion field misalignment angle and has a flat prior distribution and so is expected to be order unity, while for the Post-Inflation cosmology $\theta_i$ must be averaged, leading to $\theta_{\rm eff} = \pi/\sqrt{3}$~\cite{Hertzberg:2008wr,D'Eramo:2014rna}.

The bound Eq.~(\ref{eq:f83}) applies to the conventional cosmology with an early radiation dominated Universe and can be violated if there is a large entropy release between the GeV and MeV eras \cite{Steinhardt:1983ia}.  However, a general bound can still be derived even for this case.  An analytic estimate requiring that axions from the misaligned condensate not give more dark matter than observed today yields \cite{Kawasaki:1995vt}
\be
f_a \theta_i \simlt 10^{15} \GeV  \left( \frac{3 \MeV}{T_R} \right)^{ \scalebox{1.01}{$\frac{1}{2}$} } 
\label{eq:f95}
\ee
independent of the particle physics model for the entropy release.  $T_R$ is the reheat temperature after this entropy release, and is constrained by the $^4He$ abundance from Big Bang Nucleosynthesis (BBN) to be larger than  about 3 MeV \cite{Kawasaki:2000en}.  Thus, in both  the Pre-Inflation cosmology (with $\theta$ of order unity) or  the Post-Inflation cosmology (where $\theta_{\rm eff} = \pi/\sqrt{3}$), a low $T_R$ allows $f_a$ to be as large as $10^{15}$ GeV, which is still significantly below the scale of gauge coupling unification with TeV scale supersymmetry, characterized by the $SU(5)$ gauge boson mass $M_X \simeq 2 \times 10^{16}$ GeV.  From Eqs.~(\ref{eq:f95}), the physics of PQ breaking apparently lies well below the unified mass scale, unless $\theta_i \ll 1$.  While low $\theta_i$ could be understood in a multiverse with an anthropic requirement limiting the dark matter abundance~\cite{Linde:1987bx}, since 1983 there has been little interest in pursuing grand unified theories where PQ and unified gauge symmetries are broken together. 

The frequently quoted bound of Eq.~(\ref{eq:f83}) may not apply in supersymmetric theories.  With low energy supersymmetry the field responsible for breaking the PQ symmetry is expected to release entropy at late time possibly diluting the misalignment axions \cite{Hashimoto:1998ua}.  There is no need for introducing separate physics for axion dilution -- it can arise from the late decay of a saxion condensate.  However, whether the saxion condensate dilutes axions significantly depends on the corresponding reheat temperature and hence on $f_a$.  In any case, the bound of Eq.~(\ref{eq:f95}) still applies. Axion dilution from saxion decays has led to several studies with $f_a$ of order $10^{15} - 10^{16}$ GeV \cite{Kawasaki:2011ym}.

In this paper we study a class of Supersymmetric Grand Unified Theories where unified and PQ symmetries are broken together -- SaxiGUTs. We demonstrated that these theories are realistic and lead to a characteristic set of predictions.  The key ingredients of these theories are
\begin{itemize}
\item {\it TeV scale supersymmetry with precise gauge coupling unification.}\cite{Dimopoulos:1981yj}
The scale of the vacuum expectation value (vev) that breaks $SU(5)$ unified gauge symmetry is predicted to be 
\be
V_5 \simeq (1-4) \times 10^{16} \GeV
\label{eq:V5}
\ee
depending on threshold corrections and the value of the unified gauge coupling, $g_5$, with higher $g_5$ leading to lower $V_5$.
\item {\it Simultaneous breaking of $SU(5)$ and PQ symmetries:}
\be
V_{PQ} \, \sim \, V_5.
\label{eq:VPQ}
\ee
Such a large PQ breaking implies a Pre-Inflation cosmology.  PQ symmetry is broken before inflation and reheating after inflation does not restore the PQ symmetry.  Hence axion strings and domain walls are inflated away and are not relevant to the observable Universe.  The amount of dark matter arising from the misalignment of the axion condensate depends on the initial misalignment angle $\theta_{i}$, which we take order unity.  

\item{\it A large color anomaly of the PQ symmetry.} 
The scale $f_a$, defined by Eq.~(\ref{eq:f}), is determined by 
\be
f_a \; =  \frac{\sqrt{2} V_{PQ} }{ N_{\rm DW}} \; \simeq \; 10^{15} \GeV \left( \frac{V_{PQ}}{3 \times 10^{16} \GeV} \right) \left(  \frac{40}{N_{\rm DW}} \right),
\label{eq:f15}
\ee
where $N_{\rm DW}$ is the color anomaly of the PQ theory, also known as the domain wall number.  In SaxiGUTs it would require an accident for $N_{\rm DW}$ to be order unity, and instead we find that in typical SaxiGUTs $N_{\rm DW} \sim 10-100$.

\item {\it Doublet-triplet splitting giving a DFSZ-type axion theory.}  
A key issue in unified theories is the mass splitting between the light Higgs doublets and their $SU(5)$ partners, which must be heavy to satisfy the bound from proton decay searches.  The two Higgs doublets lie in multiplets $H$ and $\bar{H}$ of the unified theory and a large mass splitting between the weak doublets and color triplets arises from an interaction between $\bar{H} H$ and fields that break the unified symmetry.  We study the generic case where $\bar{H} H$ carries a non-zero PQ charge and hence the axion theory is of DFSZ type \cite{Dine:1981rt,Zhitnitsky:1980tq}.   This includes theories where the doublet-triplet splitting arises via fine-tuning \cite{Nilles:1981py} or via vacuum alignment~\cite{Hall:1995eq}.  
\end{itemize}

\begin{figure}
\begin{center}
\includegraphics[width=0.95\linewidth]{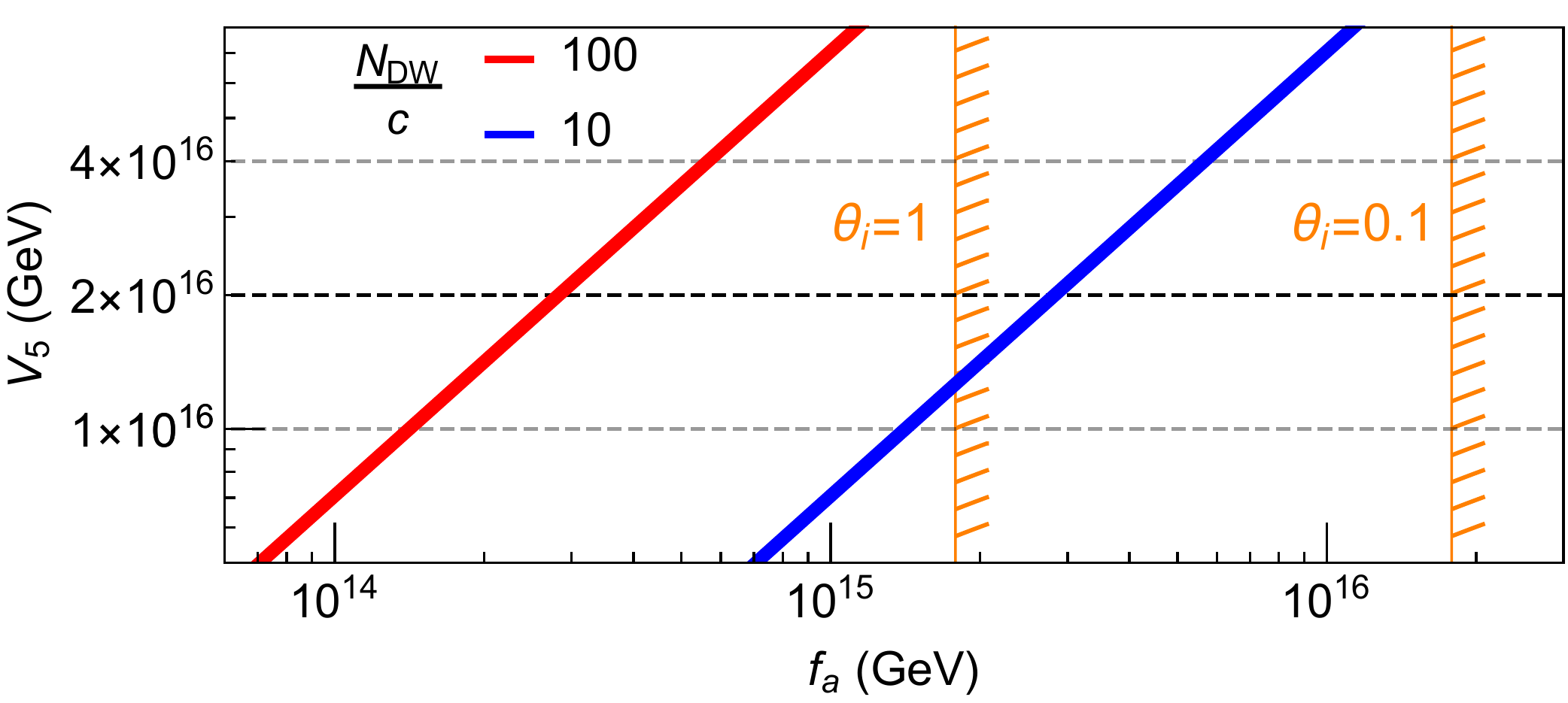}  
\end{center}
\caption{The $SU(5)$ breaking scale $V_5$ as a function of $f_a$, i.e. \Eq{eq:faV5}, for two different values of $N_{\rm DW}/c$. The orange lines show values of $f_a$, for different $\theta_i$, that yield the observed dark matter abundance for $T_R = 3$ MeV; as $T_R$ is increased, $f_a$ decreases as $1/\sqrt{T_R}$.}
\label{fig:V5fa}
\end{figure}

A large value of $N_{\rm DW}$ resolves an apparent discrepancy:  $f_a \sim 10^{15}$ GeV and $V_5 \simeq (1-4) \times 10^{16} \GeV$ become consistent with $V_{PQ} \sim V_5$. The relation between $V_5$ and $f_a$ is
\be
V_5 = \frac{N_{\rm DW}}{\sqrt{2} c} f_a \hspace{.5in} \mbox{where} \hspace{.5in} c \equiv \frac{V_{PQ}}{V_5}
\label{eq:faV5}
\ee
and $c$ is order unity.  In Fig. \ref{fig:V5fa} we illustrate this consistency between supersymmetric gauge coupling unification of Eq.~(\ref{eq:V5}) and values of $f_a$ that yield the observed dark matter from Eq.~(\ref{eq:f95}) for the two representative values $N_{\rm DW}/c = 10, 100$. 

We present our study as follows. We introduce general SaxiGUT theories, together with simple examples, in Sec.~\ref{sec:SaxiGUTs}, and in doing so we precisely define the scale of PQ breaking $V_{PQ}$, the scale of gauge coupling unification $V_5$ and the domain wall number $N_{\rm DW}$. In Sec.~\ref{sec:cosIC} we demonstrate that inflation typically generates a saxion condensate of order $V_{PQ}$ or larger. The misalignment axion DM abundance is independent of the specific SaxiGUT model and in Sec.~\ref{sec:EFT} we present an effective field theory (EFT) for the axion chiral multiplet valid below the PQ breaking scale.
We compute in Sec.~\ref{sec:saxiondilution} the dark matter abundance from axion oscillations initiated when the saxion condensate dominates the total energy of the Universe, making use of recent lattice results. We emphasize that the observed dark matter abundance calls for supersymmetry breaking at the TeV scale. The rest of the paper is devoted to additional signals for SaxiGUTs: dark radiation in Sec.~\ref{sec:DarkRad}, proton decay in Sec.~\ref{sec:pdecay}, isocurvature perturbations and CMB tensor modes in Sec.~\ref{sec:isocurvpert}.  A concise summary of correlated predictions for SaxiGUTs is provided in Sec.~\ref{sec:con}. 

\section{SaxiGUT Theories}
\label{sec:SaxiGUTs}

We study a class of theories where the grand unified symmetry $G$ and the PQ symmetry are broken by the vevs of a set of chiral superfields at the unified scale. Some of these fields carry both $G$ and PQ quantum numbers, while others may be charged under only $G$ or PQ.  In this section we first give a precise definition of the scales of vevs for PQ breaking, $V_{PQ}$, and gauge coupling unification, $V_5$. Next we give examples of SaxiGUTs and the moduli field that results from breaking PQ symmetry, and finally we discuss the domain wall number. Throughout the rest of this paper we adopt the following conventions: we denote superfields and their lowest components by boldface and regular typeface, respectively. Thus we are interested in theories of chiral superfields $\bPhi_i$ where the lowest components $\Phi_i$ get $G$ and/or PQ breaking vevs $v_i$.

\subsection{PQ and $SU(5)$ Breaking Scales}

The scale of $PQ$ breaking is defined by 
\be
V_{PQ}^2 \,=\, \sum_i q_i^2 v_i^2
\label{eq:VPQ2}
\ee
where the PQ charges $q_i$ of $\Phi_i$ are normalized such that they are all integers with $|q_i|$ as small as possible.   Gauge coupling unification depends on the spectrum of states at the unified scale that breaks $SU(5)$ symmetry ($G$ may be larger than $SU(5)$), such as the heavy SU(5) gauge bosons $X$ and Higgs colored triplets. We choose to define the symmetry breaking scale of gauge coupling unification by
\be
V_5^2 \,=\, \left( \frac{M_X}{g_5} \right)^2 \,=\, \sum_i t_i^2 v_i^2
\label{eq:V52}
\ee  
where $g_5$ is the gauge coupling evaluated at scale $V_5$.  The group theory constants $t_i^2 = (5/6,4/3)$ for $i= (24,75)$ dimensional representations are close to unity, while $t_i=0$ if $v_i$ preserves $SU(5)$.  Ignoring unified threshold corrections, gauge coupling unification implies $M_X \simeq 2 \times 10^{16}$ GeV so that $V_5 \simeq 2 \times 10^{16} \GeV /g_5$.  With the minimal matter content of the MSSM below $M_X$ the unified coupling is predicted to be $g_5 \simeq 0.7$; however, with additional matter the unified coupling could be larger.  With further uncertainty from unified threshold corrections, we adopt the range $V_5 \sim (1-4) \times 10^{16} \GeV$ for precision unification.

For low energy physics of the axion supermultiplet the key result is
\be
V_{PQ} \, = c \, V_5 \simeq c \; (1-4) \times 10^{16} \GeV.
\label{eq:VPQnum}
\ee
From Eqs.~(\ref{eq:VPQ2}) and (\ref{eq:V52}),  $c$ is typically larger than unity, although it can be less than unity if some $\Phi_i$ have $q_i = 0$.

We study DFSZ type theories where the matter and Higgs fields of the low energy theory carry PQ charges.  In particular the MSSM Higgs bilinear $\bHu \bHd$ carries non-zero PQ charge, so that the cosmological saxion condensate decays to Higgs and electroweak gauge bosons, reheating the visible sector.  Hence the $\mu$ term is generated from PQ breaking, from operators of the form $[\mathcal{G}(\bPhi_i) \bHu \bHd]_{\theta^2}$ or $[\bX \mathcal{G}(\bPhi_i, \bPhi_j^\dagger) \bHu \bHd]_{\theta^2 \bar{\theta}^2}$, where $\bX$ is a chiral superfield with a supersymmetry breaking $F$ component.

\subsection{SaxiGUT Models}

In principle the PQ symmetry could be an $R$ symmetry, as in the case of both the Nilles-Raby $SU(5)$ theory \cite{Nilles:1981py} and the Hall-Raby $SO(10)$ theory \cite{Hall:1995eq}.  However, it has been shown quite generally that in flat space supersymmetric theories with a continuous $R$ symmetry broken at scale $V_{PQ}$ the vacuum value of the superpotential is bounded by $\left| \left<W \right>\right| \leq F V_{PQ}/2$ where $F$ is the scale of supersymmetry breaking \cite{Dine:2009sw}.  Such values of $\left| \left<W \right>\right| $ are insufficient to cancel the cosmological constant in supergravity unless $V_{PQ}$ is of order the reduced Planck mass.  Hence we restrict our attention to non-$R$ symmetries.
  
Many SaxiGUTs models can be constructed as follows.  Let $\beff(\bPhi)$ be a PQ invariant product of $n$ of the $\bPhi_i$ fields, divided by $M_*^n$, where $M_*$ is the UV cutoff of the theory: $\beff = (\Pi \, \bPhi) /M_*^n$.  The superpotential takes the form
\be
\bW \, = \, \bW(\beff) + \overline{\bH} \, \bPhi_1 (1 + \beff + ...) \, \bH
\label{eq:WnonR}
\ee
where $\overline{\bH}$ and $\bH$ contain the Higgs doublets that lead to quark and charged lepton masses, and $\bPhi_1$ has PQ charge opposite to that of $\overline{\bH} \bH$. The superpotential $\bW(\beff)$ leads to a non-zero vev for $\beff$, breaking both PQ and $G$.  This could arise by introducing a singlet field $\bZ$ that drives the vev via $\bW(\beff) = M_*^2 \bZ(\beff-1)$ or from a superpotential that is a polynomial in $\beff$, $\bW(\beff) = M_*^3(\beff + \beff^2 + ...)$, where part of moduli space has $\beff$ determined to a non-zero value.  The former case requires $\beff$ to be invariant under the gauge symmetry $G$. In the latter case if $\beff$ is not gauge invariance certain terms in the polynomial expansions are absent.  Here and in Eqs.~(\ref{eq:WnonR}, \ref{eq:WSIg}) we omit the dimensionless coupling constants.  Those in the interaction involving $\overline{\bH}$ and $\bH$ are fine-tuned to yield a splitting between the doublet and triplet Higgs, leading to a TeV scale $\mu$ parameter.  Those in $\bW(\beff)$ must be chosen so that the vacuum value for $\beff$ is less than unity, allowing the PQ and gauge symmetries to be broken below the cutoff in the region of validity of the theory.

Simple $SU(5)$ theories with two $\bPhi_i$ fields include $\beff = \bSigma(+1) \bSigma(-1)/M_*^2,\; \bS(+1) \bSigma(-1)/M_*^2$ and $\beff= \bSigma(+1)^2 \bSigma(-2)/M_*^3$ where $\bS$ and $\bSigma$ are $SU(5)$ singlets and adjoints, respectively, and the PQ charges are shown in parentheses.  An example with $ \beff \sim (\bSigma_+ \bSigma_-)$ and no driver field is
\be
\bW(\bSigma_+, \bSigma_-) \, = \, M_* (\bSigma_+ \bSigma_-) + \frac{(\bSigma_+ \bSigma_-)^2}{M_*} + ... + \overline{\bH} \, \bSigma_+ \left(1 +\frac{(\bSigma_+ \bSigma_-)}{M_*^2} + ...\right) \, \bH.
\label{eq:WSIg}
\ee

Although the superpotential interactions determine the vev of $\beff$, the spontaneous breaking of the global PQ symmetry, with supersymmetry unbroken, implies a massless chiral superfield, $\bsigma(x)$, that parametrizes a degenerate moduli space of vacua.  This is lifted by the addition of soft supersymmetry breaking interactions
\be
V_{soft} \, = \, \tilde{m}^2  \left[ c_i \, \phi_i^* \phi_i  + O \left( \frac{(\phi_i^* \phi_i)^2}{M_*^2} \right) \right]
\label{eq:VnR}
\ee
where $\tilde{m}$ is the scale of supersymmetry breaking and the constants $c_i$ describe interactions between supersymmetry breaking and unified sectors.  We take $c_i >0$ so that the vacuum is determined by the quadratic terms alone.  

In the example with two fields $\bPhi_\pm$, of PQ charge $\pm q$, and $\beff = \bPhi_+ \bPhi_-/M_*^2$ determined to be $M^2/M_*^2$, the vacuum is 
\be
v_\pm = \left( \frac{c_\mp}{c_\pm} \right)^{ \scalebox{1.01}{$\frac{1}{4}$} } M 
\label{eq:vpm}
\ee
giving
\be
V_{PQ} =x q M,   \hspace{.5in} x= 
\sqrt{\left( \frac{c_+}{c_-} \right)^{ \scalebox{1.01}{$\frac{1}{2}$} } + \left( \frac{c_-}{c_+} \right)^{ \scalebox{1.01}{$\frac{1}{2}$}  }}.
\label{eq:xqM}
\ee
Small fluctuations about this vacuum are described by
\be
\bPhi_\pm \, = \, v_\pm e^{\pm q\bA/ V_{PQ}} \, ,
\label{eq:truevac}
\ee
where  $\bA$ is the canonically normalized axion chiral superfield $\bA = (s + ia)/\sqrt{2} + \theta \tilde{a}$. The saxion field $s$ is defined to be zero at the minimum of the potential and, like the axino field $\tilde{a}$, picks up a supersymmetry breaking mass, while the axion field $a$ acquires mass only from the QCD anomaly. 

The constant $c$ is easily computed in particular models.  For example, in any theory where symmetry breaking arises from SU(5) adjoints and singlets, $\bPhi_i = (\bSigma_A, \bS_\alpha)$, with PQ charges $q_i = \pm 1$
\be
c \, =  \, \frac{V_{PQ}}{V_5} = \sqrt{ \frac{6}{5} \left( 1 + \frac{\sum_\alpha v_\alpha^2}{\sum_A v_A^2} \right) }.
\label{eq:cex}
\ee
This result applies even if $G$ is larger than $SU(5)$, such as $SO(10)$, as long as the $G$ breaking vevs can be decomposed into $SU(5)$ singlets and adjoints. 

\subsection{Domain Wall Number}

A key aspect of SaxiGUTs is that the domain wall number, $N_{\rm DW}$, resulting from the field configurations of the vacuum of $\Phi_i$ is typically large.   For a non-R symmetry
\be 
N_{\rm DW} \, = \, 2 \left| \sum_a q_a T_a  \right|
\label{eq:N}
\ee
where $a$ runs over all chiral superfields of the theory, including both $\Phi_i$ and the matter and Higgs of the low energy theory.  The PQ charges  $q_a$ are normalized so that $|q_i|$ are all integers and take the smallest values possible.  $T_a$ is the Dynkin index of the color generator for representation $a$ with conventional normalization.  For $SU(5)$, $T_a = 1/2, 3/2, 5, ...$ for the $5, 10, 24, ...$ representations.  For $SO(10)$, $T_a = 1, 2, 8, 12, ...$ for the $10, 16, 45, 54, ...$ representations.  Given the size of these Dynkin indices, and that $|q_i| \ge 1$, SaxiGUTs typically have $N_{\rm DW} \gg 1$, and we will often consider the range of $N_{\rm DW}$ from 10 to 100.   

For the theories described by Eq.~(\ref{eq:WnonR}) $N_{\rm DW}$ depends on the nature of the PQ singlet function $\beff$ and on the nature of $\Phi_1$ appearing in Eq.~(\ref{eq:WnonR}). It can be summarized as $N_{\rm DW} = 2 \left| q_1 + \sum_i q_i T_i \right|$, where $i$ only runs over all $\Phi_i$.  In two field $SU(5)$ examples:  $\beff \sim \bSigma_+ \bSigma_-$ gives $N_{\rm DW}=2$ for both choices of $\Phi_1 =(\bSigma_+, \bSigma_-)$; $\beff \sim \bS \,  \bSigma$ gives $N_{\rm DW} =(8,12)$ for $\bPhi_1 =(\bS, \bSigma)$; and $\beff \sim \bSigma^2 \bSigma'$ gives $N_{\rm DW}=(8,14)$ for $\bPhi_1 =(\bSigma, \bSigma')$.  These theories have only a few small multiplets breaking the unified symmetry.  Larger unified theories will have larger $N_{\rm DW}$; for example the Hall-Raby $SO(10)$ theory has $N_{\rm DW}=94$.

\section{Initial Conditions from Inflation}
\label{sec:cosIC}

The abundance of axion dark matter is greatly affected by the decay of the saxion condensate, so it is important to study the size of this condensate, which originates from standard inflation. 
SaxiGUTs have a large domain wall number as discussed in Sec.~\ref{sec:SaxiGUTs}.  We avoid the axion domain wall and GUT monopole problems by assuming that the PQ and $G$ breaking vevs of $\Phi_i$ are non-zero during inflation so that these defects are inflated away.   
Hence, both during and after inflation we must work in the broken phase where, in the supersymmetric limit, the potential involves a complex flat direction, $\sigma(x)$, corresponding to the axion and saxion field modes.  When the potential is expanded about any point along this flat direction, the orthogonal modes have masses of order the unified scale (ignoring the possibility of further moduli).  Hence any initial values for these other modes will rapidly disappear due to damped oscillations in the expanding early Universe.  

In this section we find inflation leads to three behaviors for the initial value of the condensate $\sigma_i$ (defined with $\sigma=0$ today)
\begin{enumerate}
\item $\sigma_i \sim V_{PQ}$, where $V_{PQ}$ is the $PQ$ breaking scale today, as defined in the previous section (with $V_{PQ} \sim V_5 \sim 2 \times 10^{16}$ GeV). 
\item $\sigma_i \sim M_*$, where $M_*$ is the UV cutoff of the SaxiGUT field theory.
\item $\sigma_i = 0$.
\end{enumerate}
The third case results only in certain special situations, and is not of interest for SaxiGUTs.  The axion dark matter abundance resulting from the first two cases is computed in Sec.~\ref{sec:saxiondilution} and is found to be independent of $\sigma_i$ over the entire relevant region of parameter space.  In Sec.~\ref{sec:isocurvpert} we show that the second case allows the energy scale of inflation $E_I \sim 10^{16}$ GeV, so that tensor modes of the cosmic microwave background may be discovered at next generation experiments.

The potential for $\sigma$ requires supersymmetry breaking, and during inflation supersymmetry was broken by the physics that generated the inflaton potential $\rho_I \sim E_I^4$ \cite{Dine:1995uk}.    This physics communicates with the grand unified sector via higher-dimension operators suppressed by the cutoff scale of the theory, $M_* > V_{PQ}$, so that during inflation $\sigma$ feels a supersymmetry breaking potential
\be
V^I(\sigma) \, = \, \rho_I \left[ c_i^I \frac{\phi_i^* \phi_i}{M_*^2}  + O \left( \frac{\phi_i^* \phi_i}{M_*^2} \right)^2 \right]
\label{eq:VI}
\ee
where the constants $c_i^I$ describe interactions between inflation and unified sectors and are order unity.   The relevant supersymmetry breaking scale is $E_I^2/M_*$ and for $E_I > 10^{10}$ GeV this scale dominates that from the usual soft supersymmetry breaking interactions, $\tilde{m} \sim$ TeV.  

For $c_i^I>0$, only the leading term in $1/M_*^2$ in Eq.~(\ref{eq:VI}) is needed to determine $\sigma$ and the resulting potential leads to a minimum for the real component of $\sigma$ which acquires a mass of order of $E_I^2/M_*$. For any cutoff $M_*$ less than the reduced Planck scale $M_{Pl} = 2.4 \times 10^{18}$ GeV, the saxion field will undergo oscillations during inflation, since $m_{sI} > 3H_I$, and will settle to the minimum of this potential.  Thus inflation determines an initial value for the saxion field.  
 
In the SaxiGUT models discussed in Sec.~\ref{sec:SaxiGUTs}, the superpotential constrains a product of $n$ fields, called $f(\phi_i)$, to some value $M^n$, with $M$ of order $V_{PQ}$. 
With $c_i^I$ of order unity, at the minimum $\phi_i = v_i^I$ take values of order $M$.  For example, in the case of $f \sim \Phi_+ \Phi_-$  
\be
v^I_\pm = \left( \frac{c^I_\mp}{c^I_\pm} \right)^{ \scalebox{1.01}{$\frac{1}{4}$}  } M.
\label{eq:vpmI}
\ee
Comparing with Eqs.~(\ref{eq:vpm}, \ref{eq:xqM})
\be
 \sigma_i = (x^I-x) \; q M
 \label{eq:sigmai}
 \ee
 where $x^I$ is the same function of $c^I_\pm$ as $x$ is of $c_\pm$.  For generic order unity parameters $(c^I_\pm, c_\pm)$, $\sigma_i \sim V_{PQ}$ so that theories with $c^I_\pm>0$ are of the first type listed above.


In certain theories moduli space may have special symmetry points.  If the interactions leading to the symmetry breaking potentials respect these discrete symmetries, then the constants $c_i$ and $c_i^I$ may be equal to the symmetry point values.  With $f \sim \Phi_+ \Phi_-$, imposing a $Z_2$ symmetry $\Phi_+ \leftrightarrow \Phi_-$ leads to $c_+ = c_-$ and $c_+^I = c_-^I$.  Under these circumstances the minimum of the potential is at the symmetry point both during and after inflation so that $x^I = x$ and there is no condensate $\sigma_i = 0$, giving the third case listed above.

If one or more of $c_i^I$ are negative, the quadratic term of Eq.~(\ref{eq:VI}) leads to field values much greater than $V_{PQ}$.  Positive higher order terms are required to stop runaway behavior, and in this case the generic expectation is that $\sigma_i \sim M_*$, the second case listed above.

The role of inflation in the above discussion is to determine $\sigma_i$.  However, even if $E_I \ll 10^{10}$ GeV, so the inflationary era does not determine $\sigma_i$, there is no reason why this initial value should be within $V_{PQ}$ of the minimum determined by Eq.~(\ref{eq:VnR}).

\section{The Effective Theory Below The Unified Scale}
\label{sec:EFT}

To describe axion and saxion physics below the PQ breaking scale, we take a model-independent approach and write down an EFT.  This is sufficient for the computation in the next section of the axion dark matter abundance, which is therefore independent of many model-dependent features of SaxiGUTs.  Since $V_{PQ}$ is much larger than the SUSY breaking scale, we construct a supersymmetric EFT where the PQ symmetry is non-linearly realized~\cite{Zumino:1979et,Bellazzini:2011et}. In this section we write down such an EFT and give the partial widths for saxion decays. Details about the EFT and calculations can be found in Appendix \ref{app:saxion}.

Our conceptual starting point is a UV complete theory where the PQ symmetry is broken by the vevs $v_i$ of chiral superfields $\bPhi_i$. We assign to $\bPhi_i$ a PQ charge $q_i$, such that a PQ rotation with angle $\alpha$ in the UV theory induces  
\be
\bPhi_i \; \rightarrow \; \exp\left[ i q_i \alpha \right] \bPhi_i  \ .
\ee
At energies below the PQ breaking scale we have a massless Goldstone superfield 
\be
\bA = \frac{s + i \, a}{\sqrt{2}} + \sqrt{2} \theta \tilde{a} + \theta^2 F_A \ .
\label{eq:aexpansion}
\ee
The different components of $\bA$ are massless, since the axion $a$ mass is protected by being a Goldstone boson and the degeneracy in the multiplet is ensured by SUSY.  The breaking of SUSY will provide masses for the saxion $s$ and the axino\footnote{We do not consider axino LSPs. Axinos can be copiously produced through Freeze-In~\cite{Cheung:2011mg}, and if they are the LSP, their cosmic density could be depleted by a dilution mechanism analogous to the one discussed in this work~\cite{Co:2015pka}.} $\tilde{a}$, but not for the axion. In the low-energy EFT the PQ symmetry is nonlinearly realized 
\be
\bA \; \rightarrow \; \bA + i \, \alpha \, V_{PQ} \ ,
\label{eq:bAshift}
\ee
where the effective scale of PQ breaking $V_{PQ}$ was introduced in \Eq{eq:VPQ2}. The axion superfield $\bA$ is the low-energy degree of freedom associated with the PQ breaking fields in the UV, which can be expanded around their vevs as follows
\be
\bPhi_i = v_i \exp\left[ q_i \frac{\bA}{V_{PQ}} \right] \ .
\label{eq:bPhi}
\ee

The EFT Lagrangian and the details of the saxion decay widths calculation are presented in App.~\ref{app:saxion}. Here, we only report the results relevant to the discussion in Sec.~\ref{sec:saxiondilution}. The saxion has two possible decay channels. K\"ahler potential interactions induce saxion decays to two axions through the operator
\begin{align}
\label{eq:Lsaa} \mathcal{L}_{s a a} = & \, \frac{\kappa}{\sqrt{2} \, V_{PQ}} \, s \, \partial^\mu a \partial_\mu a \ , \\
\label{eq:kappadef} \kappa \equiv & \, \sum \frac{q_i^3  \, v_i^2}{V_{PQ}^2} \ .
\end{align}
For models with only a single PQ breaking field, or theories with more than one but all with the same PQ charge, we have $\kappa = 1$. In more general cases $\kappa$ is a free parameter. The associated decay width results in
\be
\Gamma_{s \rightarrow a a} = \  \frac{ \kappa^2 }{64\pi} \frac{m_s^3}{V_{PQ}^2} \ .
\label{eq:Gammasaa}
\ee

Superpotential and SUSY breaking interactions are responsible for visible saxion decays to Higgs bosons through the operators
\be
V_{s H_u H_d} = \sqrt{2} \, q_\mu  \frac{\mu^2}{V_{PQ}} \, s \left(H_u^\dag H_u + H_d^\dag H_d  \right) +
q_\mu \,  \frac{m_A^2 \, \sin2\beta}{V_{PQ}}  \frac{s}{2 \sqrt{2}} \left( H_u H_d  + {\rm h.c.}  \right) \ .
\label{eq:sHH}
\ee
Here, $q_\mu$ is the PQ charge of the MSSM $\mu$ term, $\tan\beta$ is the ratio of the Higgs vevs and $m_A$ is the mass of the CP-odd Higgs boson $A$. All the partial decay widths in the decoupling limit ($m_A \gg m_Z$) can be found in App.~\ref{app:saxion}. In this work we use the large $\tan\beta$ limit of those expressions 
\be
\Gamma_{s \, \rightarrow \, {\rm visible}} = \mathcal{D} \; \times \; \frac{q^2_\mu \mu^4}{16 \pi m_s V_{PQ}^2} \ ,
\label{eq:GammaSaxionVisible}
\ee 
where the overall multiplicative factor $\mathcal{D}$ counts the number of final states kinematically allowed. For decays to SM final states only (Higgs boson and longitudinal weak bosons) we have $\mathcal{D} = 4$. If decays to heavier Higgs bosons ($H$, $A$ and $H^\pm$) are also accessible then $\mathcal{D} = 8$.  The result (\ref{eq:GammaSaxionVisible}) is valid even if the TeV scale theory is larger than the MSSM, provided that the interactions responsible for saxion decays are dominated by the $\mu$ term.

\section{Axion Dark Matter and Dilution from Saxions}
\label{sec:saxiondilution}

\begin{figure}
\begin{center}
\includegraphics[width=1\linewidth]{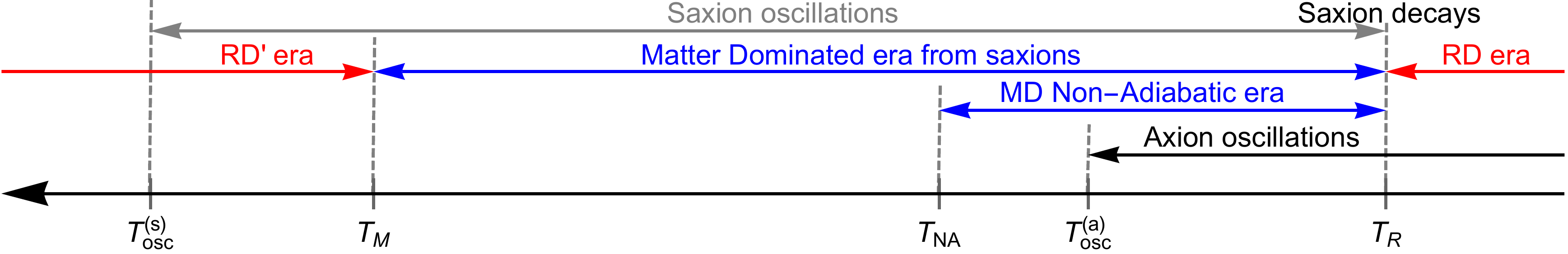}
\end{center}
\caption{A summary of different eras in saxion cosmology.}
\label{fig:SaxionCosmology}
\end{figure}

In SaxiGUT theories the unified and PQ symmetries are broken before inflation.  In the absence of supersymmetry breaking there is a complex flat direction corresponding to the axion and saxion modes.    During inflation the vacuum energy of the inflaton field breaks supersymmetry, generating a potential for the saxion field.  In Sec.~\ref{sec:cosIC} we argue that this generically leads to an initial displacement of the saxion field around today's minimum of $\sigma_i \sim V_{PQ}$. This condensate decays at a rate given by \Eq{eq:GammaSaxionVisible} and, given the large value of $V_{PQ}$, the decays occur after the QCD phase transition, diluting the axion density. This invalidates the usual cosmological bound on $f_a$ of Eq.~(\ref{eq:f83}) and in this section we compute the axion dark matter abundance in SaxiGUTs. We begin by summarizing the relevant cosmological evolution, illustrated in Fig. \ref{fig:SaxionCosmology}.

After inflation the saxion field remains fixed at $\sigma_i$ due to Hubble friction until the Hubble parameter drops to $3 H \sim m_s$ when it starts to oscillate at temperature 
\be
\label{eq:SaxTosc}
T_{osc}^{(s)} = \left(  \frac{10}{\pi^2g_*(T_{osc}^{(s)})}  \right)^{\scalebox{1.01}{$\frac{1}{4}$} } \sqrt{m_s M_{Pl}} ,
\ee
where $g_*(T)$ is the effective number of degrees of freedom in the thermal bath at temperature $T$. This happens during an early Radiation Dominated era (RD$'$). Once the saxion field starts oscillating, its energy density red-shifts as non-relativistic matter and thus decreases as $a^{-3}$, where $a$ is the scale factor of the FRW metric. It eventually dominates over radiation, and the Universe enters an early matter-dominated (MD) era at temperature $T_M$ where the saxion and radiation energy densities are equal
\be
m_s^2 \sigma_i^2 \left( \frac{T_M}{T_{osc}^{(s)}} \right)^3 = \frac{ \pi^2}{30} g_*(T_M) \ T_M^4 \ .
\ee
Assuming $g_*(T_{osc}^{(s)}) \approx g_*(T_M)$, we find
\be
\label{eq:SaxTM}
T_M =   3  \left( \frac{10}{g_*(T_M) \pi^2}  \right)^{\scalebox{1.01}{$\frac{1}{4}$}} \frac{m_s^{1/2} \sigma_i^2}{M_{Pl}^{3/2}} \ .
\ee
This MD era consists of two phases -- adiabatic (MD$_A$) and non-adiabatic (MD$_{NA}$), as detailed in Ref. \cite{Co:2015pka}.   During the MD$_A$ phase, 
the radiation energy density is dominated by the red-shifted initial radiation, with the saxion decay products providing a sub-dominant contribution. However, the relativistic decay products of the saxion eventually become the dominant form of radiation at temperature 
\be
\begin{split}
\label{eq:SaxTNA}
T_{NA} = & \left(\frac{ \ 3^{3} \ 5^{5/4} }{2^{27/4} \pi ^{9/2}} \frac{q_\mu^{4} \mathcal{D}^{2} \mu^8 \sigma_i^{2} M_{Pl}^{1/2} }{m_s^{3/2} V_{PQ}^{4} \ g_*(T_R) \ g_*(T_M)^{1/4}} \right)^{ \scalebox{1.01}{$\frac{1}{5}$} } \\ 
\simeq  & \ 0.2 \, \text{GeV} \, q_\mu^{4/5}  \left(\frac{\mathcal{D}}{4} \right)^{ \scalebox{1.01}{$\frac{2}{5}$} } \left( \frac{\mu}{3 \text{ TeV}} \right)^{ \scalebox{1.01}{$\frac{13}{10}$} } \left( \frac{\mu}{m_s} \right)^{ \scalebox{1.01}{$\frac{3}{10}$} }  \left( \frac{\sigma_i}{V_{PQ}} \right)^{ \scalebox{1.01}{$\frac{2}{5}$} } \left( \frac{2 \times 10^{16} \ \text{GeV}}{V_{PQ}} \right)^{ \scalebox{1.01}{$\frac{2}{5}$} }   \ ,
\end{split}
\ee
where we used $g_*(T_R) = 10.75$ and $g_*(T_M) = 228.75$ for the full SM and MSSM values. At temperatures below $T_{NA}$, saxion decays reheat the Universe and a large amount of entropy is released. Finally, most of the saxions decay when $H \sim \Gamma_s$ at the reheat temperature
\be
\label{eq:SaxTR}
T_R  \, =  \,   \left(  \frac{90}{\pi^2g_*(T_R)}  \right)^{ \scalebox{1.01}{$\frac{1}{4}$} }  \sqrt{\Gamma_s M_{Pl}} \, \simeq  \,   \left(  \frac{90}{\pi^2g_*(T_R)}  \right)^{ \scalebox{1.01}{$\frac{1}{4}$} }  \frac{q_\mu \sqrt{\mathcal{D}}}{4\sqrt{\pi}} \frac{\mu^2 M_{Pl}^{1/2}}{V_{PQ} \sqrt{m_s} } \, ,
\ee
where $\mathcal{D}$ denotes the number of final states kinematically accessible in saxion decay and in the second expression we recall that $\Gamma_s$ is dominated by $\Gamma_{s \, \rightarrow \, {\rm visible}}$.  This reheat temperature plays a key role in our analysis of the cosmological axion abundance and its dependence on $\mu$, as well as other dimensionless parameters, is shown in the right panel of Fig. \ref{fig:MD_NACondition}.   It is remarkable that the relevant range of $T_R$ that leads to the observed dark matter results for $\mu$ of order $(1-10)$ TeV.  After saxion decay, the Universe returns to a Radiation Dominated era (RD). 
The saxion cosmology described in this section is summarized in Fig. \ref{fig:SaxionCosmology}.

\subsection{Analytic Results}
\label{sec:AxionAnalytic}

The cosmological axion abundance depends sensitively on the turn-on of axion field oscillations.  In SaxiGUTs this occurs during the saxion MD era at a lower temperature than in the usual RD case.
We assume the temperature dependence of the axion mass takes the form
\begin{align}
\label{eq:maT}
m_a(T) = &
\begin{cases} 
      \, m_a(0)  \left(\frac{\Lambda}{T} \right)^{n} & T \ge \Lambda \\
      \, m_a(0)   & T \le \Lambda
   \end{cases}  \\
   \label{eq:ma}
m_a(0) = & \, 6 \, \text{eV} \left( \frac{10^6 \text{GeV}}{f_a} \right) ,
\end{align}
where $\Lambda$ is the QCD phase transition scale. A lattice calculation in the quenched approximation gives $2n \simeq 6.8$ \cite{Borsanyi:2015cka}, in good agreement with the dilute instanton gas approximation.  However, including light dynamical quarks leads to a much lower value, $2n = 2.7$ \cite{Bonati:2015vqz}.  We show results for both results and take $\Lambda = 150$ MeV.   These lattice results have a limited domain of validity in temperature, but are sufficient for axion oscillations in the MD era.

We calculate the axion abundance from the misalignment mechanism and saxion dilution following Ref.~\cite{Kawasaki:1995vt}. The calculation is simplified by employing the adiabatic condition that, despite the varying axion mass, the number densities of both the axion and saxion scale the same way as $a^{-3}$ after the axion starts to oscillate at temperature $T_{osc}^{(a)}$. For convenience, we define $\xi \equiv m_a(T_{osc}^{(a)})/m_a(T_R) \le 1$. If $T_{osc}^{(a)} \le \Lambda$, $\xi = 1$; otherwise, $\xi = (\Lambda/T_{osc}^{(a)})^n$. We obtain the axion abundance by
\be
\label{eq:AxionAbundance}
\left. \frac{\rho_{a}}{s}  \right|_{T_R} = \left. \frac{3}{4}  \frac{\rho_{a}}{\rho_s}  \right|_{T_{R}}  T_R = \left.  \frac{3}{4} \frac{\rho_{a}}{\rho_s} \right|_{T_{osc}^{(a)}} \frac{T_R}{\xi}    = \frac{9}{8}  \frac{f_a^2 \theta_{i}^2}{ M_{Pl}^2}  \frac{T_R}{\xi} ,
\ee
where we have taken into account the varying axion mass by $\xi$ and used $\rho_a(T_{osc}^{(a)}) = m_a^2(T_{osc}^{(a)}) f_a^2 \theta_{i}^2/2$ as well as the saxion energy density $\rho_s(T_{osc}^{(a)}) = m_a^2(T_{osc}^{(a)}) M_{Pl}^2/3$. Setting the axion abundance equal to that of the observed DM gives
\begin{align}
\label{eq:Teq}
T_e =  \, & \frac{3}{2} \dfrac{f_a^2 \theta_{i}^2 T_R}{ M_{Pl}^2 \xi} \\
\label{eq:Omegah2}
\Omega_a h^2 = \, & 0.12 \ \theta_{i}^2 \ \xi^{-1} \left( \frac{f_a}{9\times10^{14} \, \text{GeV}} \right)^2  \left( \frac{T_R}{3 \, \text{MeV}} \right) 
\end{align}
where $T_e \approx 0.6 \text{ eV}$ is the usual temperature of matter radiation equality. 

To find $T_{osc}^{(a)}$, one can make use of the temperature scaling of the saxion energy density during the MD$_{\text{NA}}$ era, $\rho_s \propto T^8$ \cite{Co:2015pka}, and of the Hubble constant $3H \sim  m_a(T_{osc}^{(a)})$ at the time of axion oscillation as well as $H \sim T_R^2/M_{Pl}$ at the end of reheating
\be
\label{eq:findTosc}
\frac{\rho_s(T_{osc}^{(a)})}{\rho_s(T_R)} = \left( \frac{T_{osc}^{(a)}}{ T_R} \right)^8 = \frac{ m_a^2(T_{osc}^{(a)}) M_{Pl}^2/3}{ \frac{\pi^2}{30} g_*(T_R) T_R^4}  
\ee
\be
\label{eq:Tosca}
T_{osc}^{(a)} =  \, \left( \frac{\sqrt{10}}{\pi}   \frac{m_a(0) \Lambda^{n} M_{Pl}T_R^2}{\sqrt{g_*(T_R)}} \right)^{ \scalebox{1.01}{$\frac{1}{4+n}$} } \ \ \ \ \ \ \ \ \text{for} \ \ T_{osc}^{(a)}  \ge \Lambda ,
\ee
and the results for $T_{osc}^{(a)}  \le \Lambda$ can be easily obtained by setting $n=0$. This can be used to show that the particular choice of parameters in Eq.~(\ref{eq:Omegah2}) corresponds to $ T_{osc}^{(a)} < \Lambda$ and thus $\xi=1$. From Eq.~(\ref{eq:Tosca}), we can find the condition for $T_{osc}^{(a)} = \Lambda$ gives a critical value for $T_R$
\be
T_{R}^{(c)} = 10 \, \text{MeV} \left(\frac{\Lambda}{150 \, \text{MeV}}\right)^2   \left(\frac{ f_a }{10^{15} \, \text{GeV}}\right)^{\scalebox{1.01}{$\frac{1}{2}$} } \left(\frac{ g_*(T_R) }{10 }\right)^{ \scalebox{1.01}{$\frac{1}{4}$} }, 
\ee
shown as the blue dotted lines in Fig. \ref{fig:MisalignContours}. This demonstrates that, for the range of $T_R$ of interest, the axion can oscillate before or after QCD phase transition; in the former case, the use of Eq.~(\ref{eq:maT}) is needed. The axion oscillation temperature in Eq.~(\ref{eq:Tosca}) leads to
\begin{align}
\label{xi}
\xi^{-1} = \, & \left(\frac{\sqrt{10}}{\pi}  \dfrac{m_a(0) M_{Pl} T_R^2}{\Lambda^4 \sqrt{g_*(T_R)}}  \right)^{\tfrac{n}{4+n}}    \ \ \ \ \ \ \ \ \text{for} \ \    T_{osc}^{(a)} \ge \Lambda 
\end{align}
which together with \Eq{eq:Omegah2} gives the axion dark matter abundance for any scenario.

\begin{figure}
\begin{center}
\includegraphics[width=0.495\textwidth]{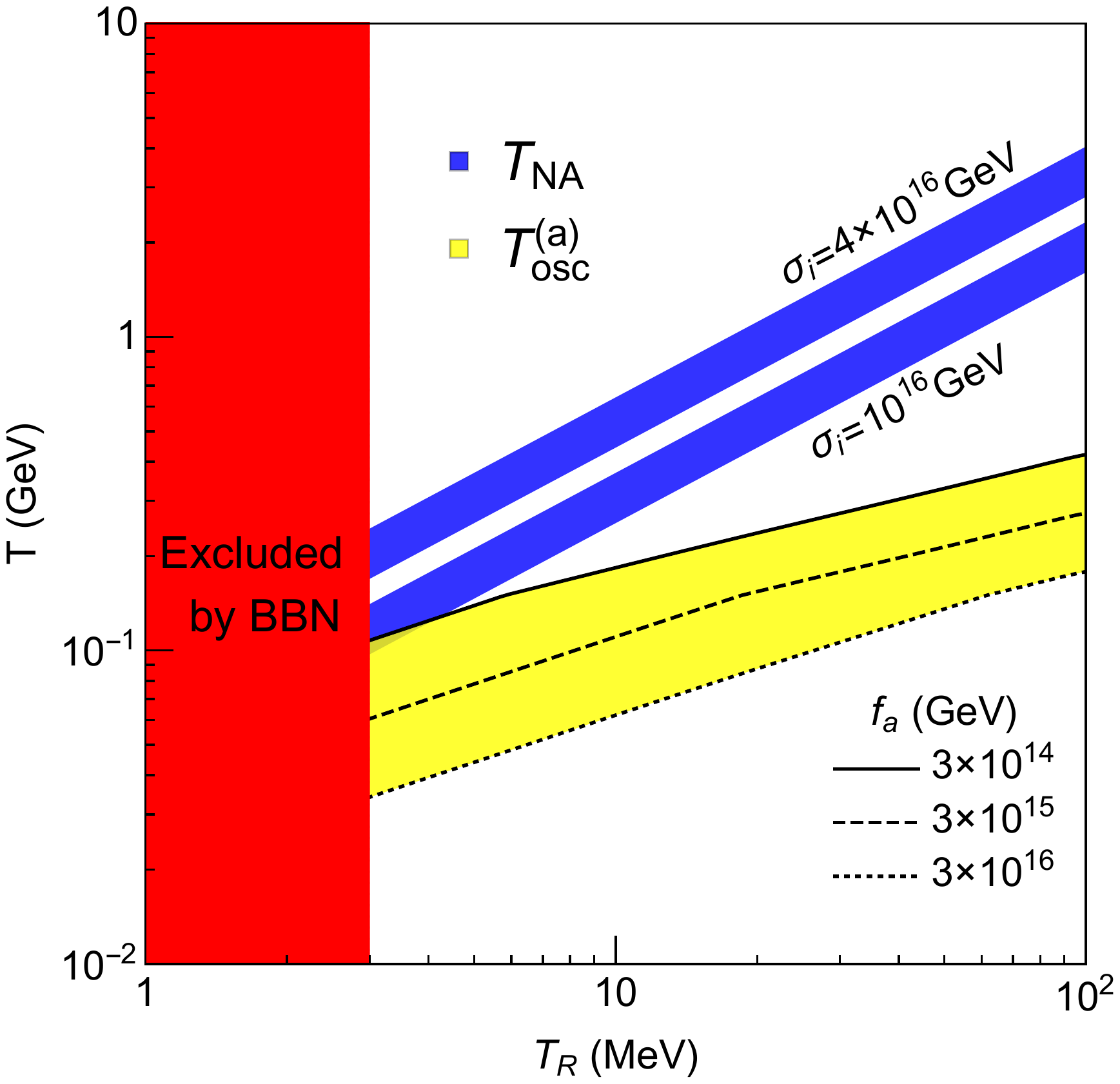} 
\includegraphics[width=0.495\textwidth]{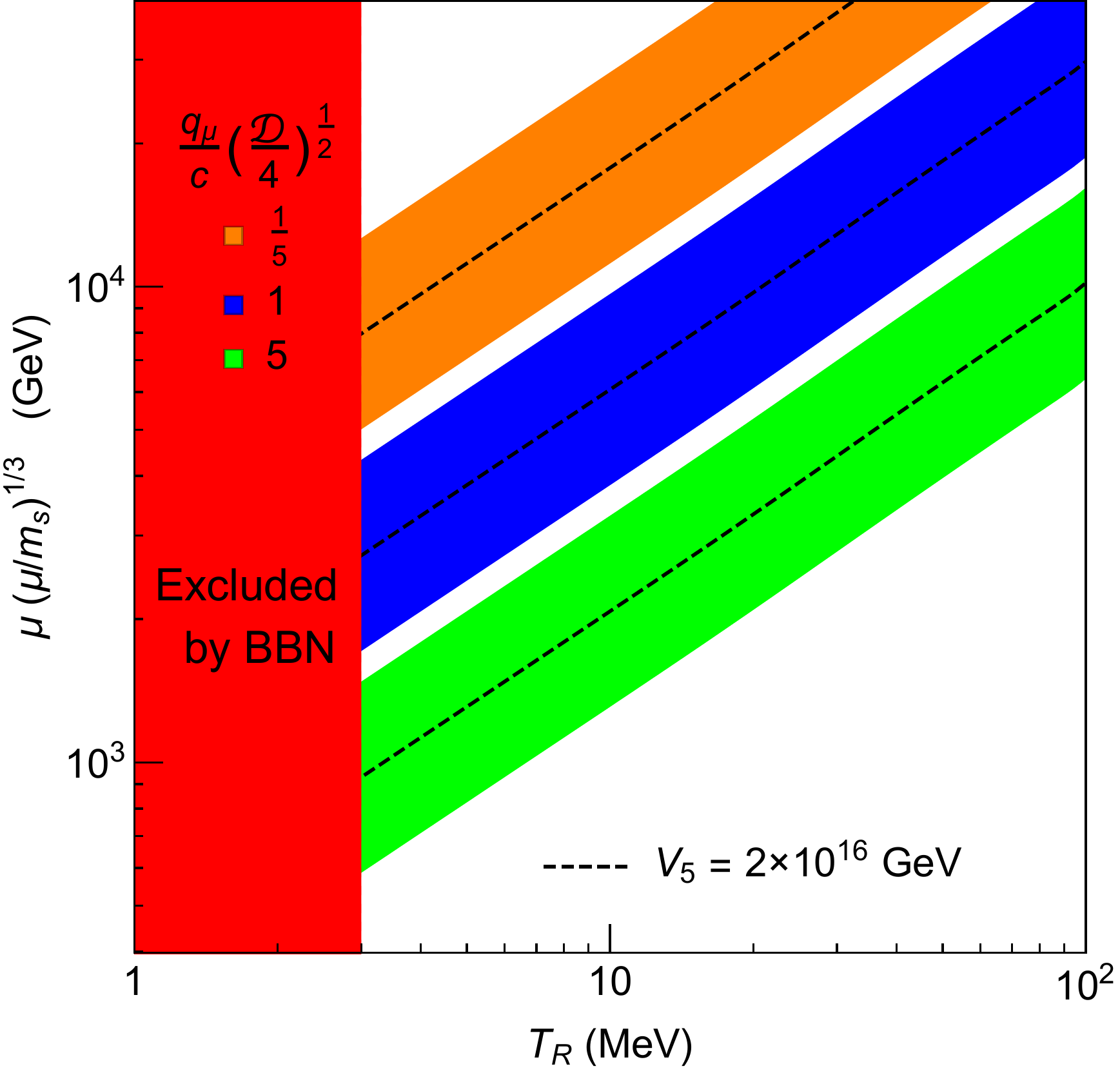}
\end{center}
\caption{\textbf{Left Panel:} the two blue bands represent $T_{NA}$ with $m_s$ varying from 250 GeV to 10 TeV, each with a different choice of $\sigma_i$. The yellow band is the possible range of the axion oscillation temperature with black lines referring to different $f_a$ used in the calculation. The red shaded region is excluded by BBN. \textbf{Right Panel:} The value of $\mu (\mu/m_s)^{1/3}$ needed for a given $T_R$ using \Eq{eq:SaxTR} with different values of $\left(q_\mu/c \right) \sqrt{\mathcal{D}/4}$. The black dashed lines are for $V_5=2\times10^{16}$ GeV while the color bands are created from varying $V_5$ by a factor of 2 each way.}
\label{fig:MD_NACondition}
\end{figure}

In the analytic and numerical analyses, we have assumed $T_{NA} > T_{osc}^{(a)}$ so that axions oscillate in the MD$_{\text{NA}}$ era. For a choice of $T_R$, $\sigma_i$ and $m_s$, $T_{NA} = (T_M T_R^4)^{1/5}$ \cite{Co:2015pka} can be calculated using \Eq{eq:SaxTM}. Similarly, for a given $T_R$ and $f_a$, one can find $T_{osc}^{(a)}$ from Eq.~(\ref{eq:Tosca}). These results for $T_{NA}$ and $T_{osc}^{(a)}$ are shown in Fig.~\ref{fig:MD_NACondition}, where the blue band is $T_{NA}$ with $m_s$ varying from 250 GeV to 10 TeV, each for a difference value of $\sigma_i$, while the yellow band is $T_{osc}^{(a)}$ with $f_a$ ranging from $3\times10^{14}$ GeV to $3\times10^{16}$ GeV. Figure~\ref{fig:MD_NACondition} clearly justifies the assumption that the axion always starts to oscillate during the MD$_{NA}$ era.    Figure~\ref{fig:MD_NACondition} also shows that $T_{osc}^{(a)}$ is sufficiently small for the validity of the lattice results of Refs.~\cite{Borsanyi:2015cka} and \cite{Bonati:2015vqz}.

\subsection{Numerical Analysis and Results}
\label{sec:FullNumerical}

We begin by giving the equations for the evolution of the cosmological background, of radiation and matter, during the epoch where axion field oscillations turn on and undergo dilution.  We include the temperature dependence of $g_*(T)$ which is strong near the QCD phase transition.

The starting point is entropy production from the matter decay
\begin{align}
dS = \, d(s a^3) = d\left(\frac{2\pi^2}{45} g_*(T) T^3 a^3 \right) & =  \,  \frac{dQ}{T} = \frac{\rho_M \Gamma_M a^3 dt}{T} 
\end{align}
and by the chain rule we obtain
\be
\label{eq:TempPDE}
\frac{2\pi^2}{45} T^3 \left( \frac{dg_*(T)}{dT}  T   + 3 g_*(T)  \right) \frac{dT}{dt}  =   \, \rho_M \Gamma_M - \frac{2\pi^2}{15}  g_*(T) T^4 H.
\ee
The matter energy density $\rho_M$ evolves according to
\be 
\label{eq:Hubble_gstar}
\dot{\rho}_M + 3 H \rho_M  = - \Gamma_M \rho_M ,\\
\ee
where $\Gamma_M$ is the matter decay rate, and the Friedman equation is
\be
\label{eq:Hubble}
H  = \frac{\sqrt{ \rho_M +  \frac{\pi^2}{30} g_*(T) T^4}}{ \sqrt{3}M_{Pl}} .
\ee
In the case of constant $g_*$, Eq.~(\ref{eq:TempPDE}) reduces to $\dot{\rho}_R + 4 H \rho_R  = \Gamma_M \rho_M$ where $\rho_R = \frac{\pi^2}{30} g_*(T) T^4$ is the radiation energy density. In our numerical calculation, $g_*(T)$ is computed using the masses of the SM particles and of the SUSY particles, assumed degenerate at 1 TeV. The initial condition for $\rho_M$ is set at some high temperature $T_i = 10^{16} \text{ GeV}$, which does not affect the axion dark matter density as long as $T_{osc}^{(a)} < T_{NA}$ (see Ref.~\cite{Co:2015pka}),
\be
\rho_{Mi} =  \frac{\pi^2}{30} g_*(T_M) \, T_M \, T_i^3
\ee
where $T_M$ is defined in Eq.~(\ref{eq:SaxTM}) and $g_*(T_i) = g_*(T_M) = 228.75$. The majority of the saxions decay when the Hubble constant is comparable to its decay rate. Since $\Gamma_M \simeq  \Gamma_{s \rightarrow \text{visible}}$, we can trade $\Gamma_M$ for the reheat temperature $T_R$ using Eq.~(\ref{eq:SaxTR}).  The above equations allow a numerical evaluation of the cosmological background, in particular of $H(T)$.

\begin{figure}[!ht]
\begin{center}
\includegraphics[width=0.495\textwidth]{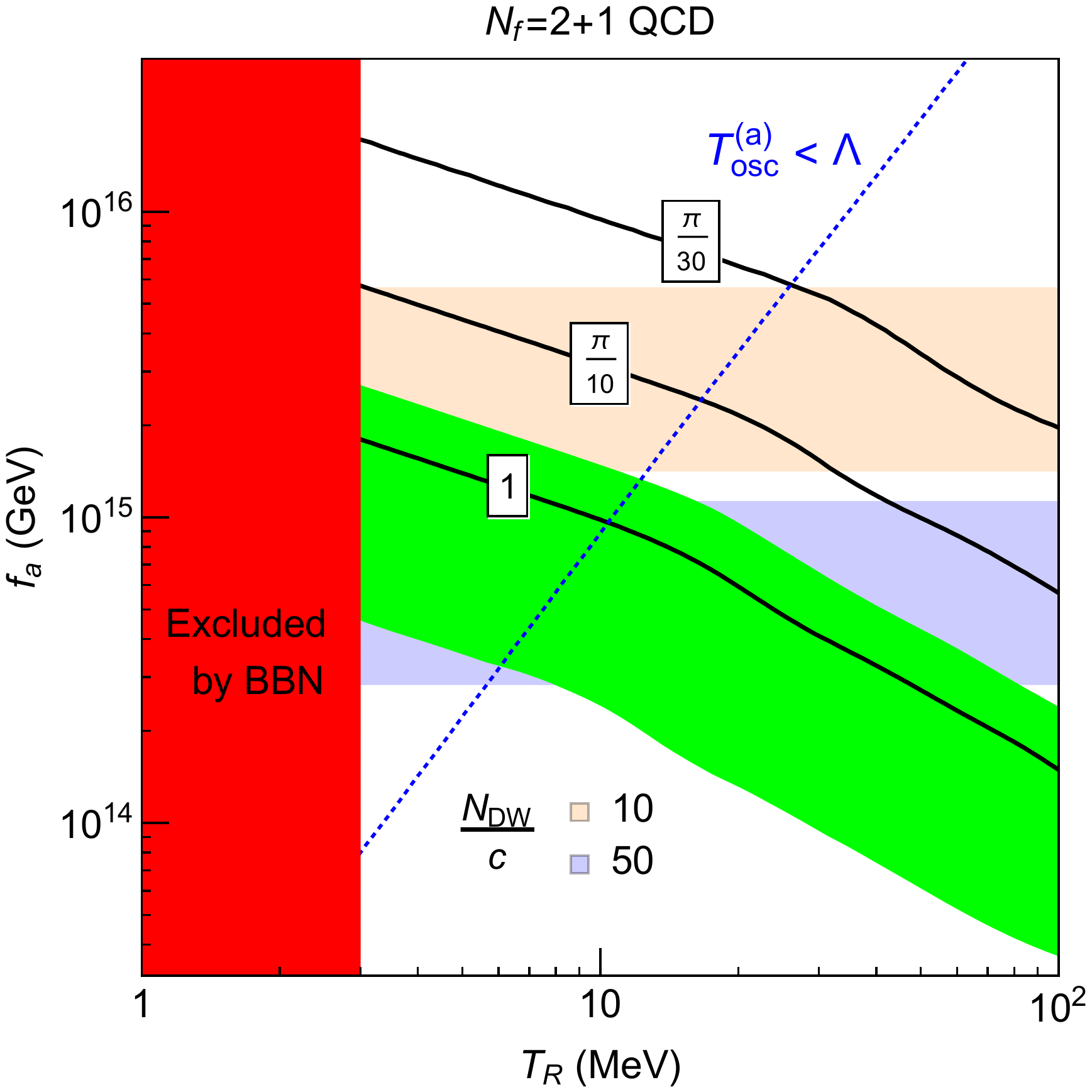}
\includegraphics[width=0.495\textwidth]{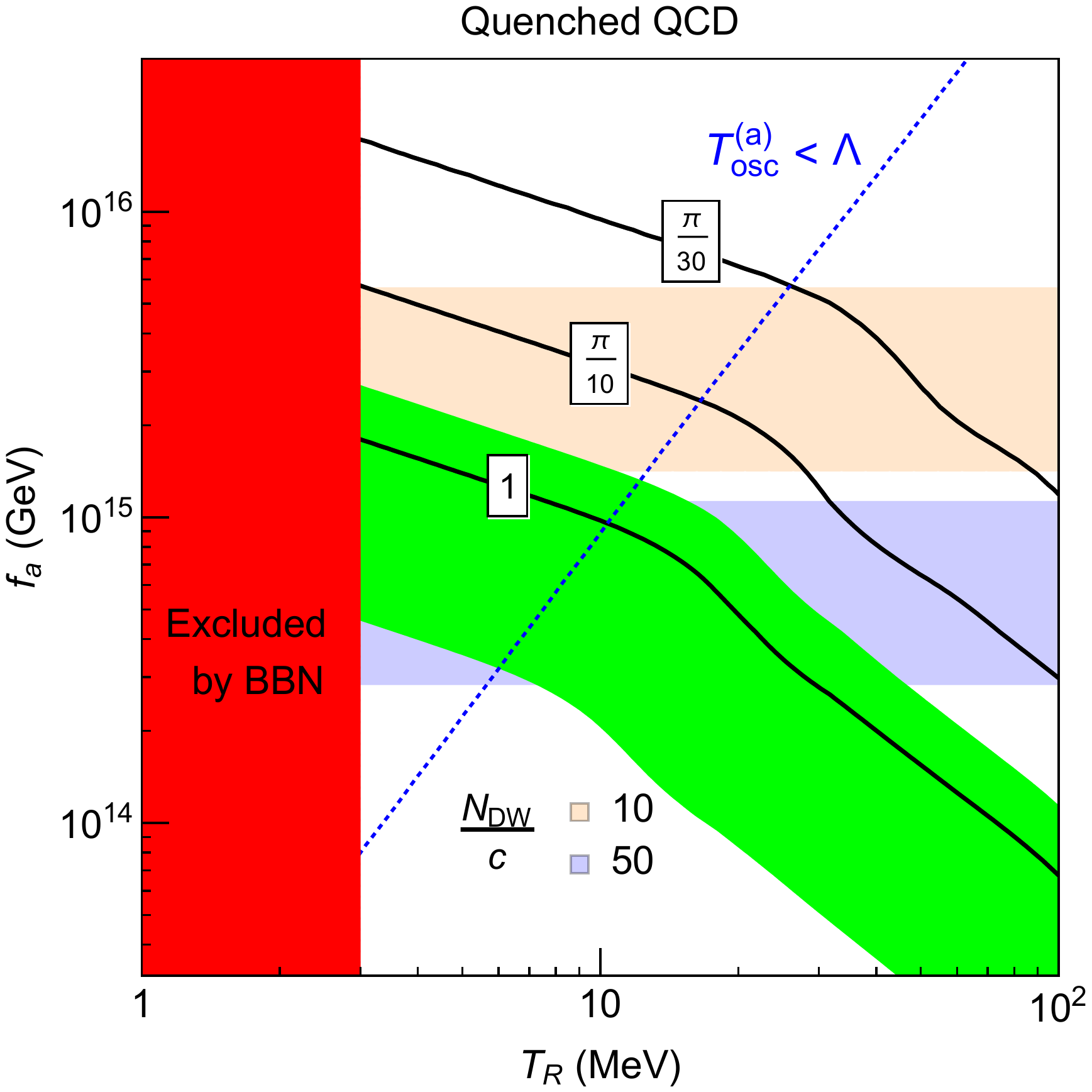}
\end{center}
\caption{Contour lines of the misalignment angle that give the observed dark matter abundance; the left (right) panel is for $2n=2.7 \ (6.8)$ for \Eq{eq:maT}. The green band refers to the $1\sigma$ range of $\theta_{i}$ of Eq.~(\ref{eq:ThetaOneSigma}). The orange (blue) band, for $N_{\rm DW}/c$ = 10 (50), shows the range of $f_a$ corresponding to $V_5=(1 - 4) \times 10^{16}$ GeV.}
\label{fig:MisalignContours}
\end{figure}

The axion field oscillation and energy density equations are
\begin{align}
\label{eq:AxionEOM}
\ddot{\varphi} + 3 H \dot{\varphi} & = -m_a^2 f_a \sin( \varphi /f_a) \\
\label{eq:AxionEnergy}
\rho_a(t) & = \frac{1}{2} \left(m_a^2(t) \, \varphi^2(t) + \dot{\varphi}^2(t) \right) 
\end{align}
with $H(T)$ determined above and $m_a(T)$ given in \Eq{eq:ma}.  These are evolved from initial conditions
$\varphi_i =  \, f_a \, \theta_{i}$ and $\dot{\varphi}_i = 0$.  Since the axion starts to oscillate in the MD$_{NA}$ era, $T_M$ is irrelevant \cite{Co:2015pka}. Furthermore, \Eq{eq:SaxTM} implies that the saxion initial oscillation amplitude $\sigma_i$ does not affect the axion oscillation either. As a result, in this calculation, the free parameters are $T_R$, $f_a$, and, $\theta_{i}$. Requiring the axion abundance to be equal to the observed dark matter abundance
\be
\left. \frac{\rho_{a}}{s}  \right|_{T_R}  = T_e ,
\ee
determines the misalignment angle for each given $(T_R, f_a)$, giving the contour plot of Fig.~\ref{fig:MisalignContours}. For $\theta_{i}$ of order unity, $f_a$ can be as high as $ 2 \times 10^{15}$ GeV without upsetting the BBN bound, which then allows $V_{PQ}$ to be as high as the GUT scale with $N_{\rm DW} \sim 10$.

\begin{figure}
\begin{center}
\includegraphics[width=0.495\textwidth]{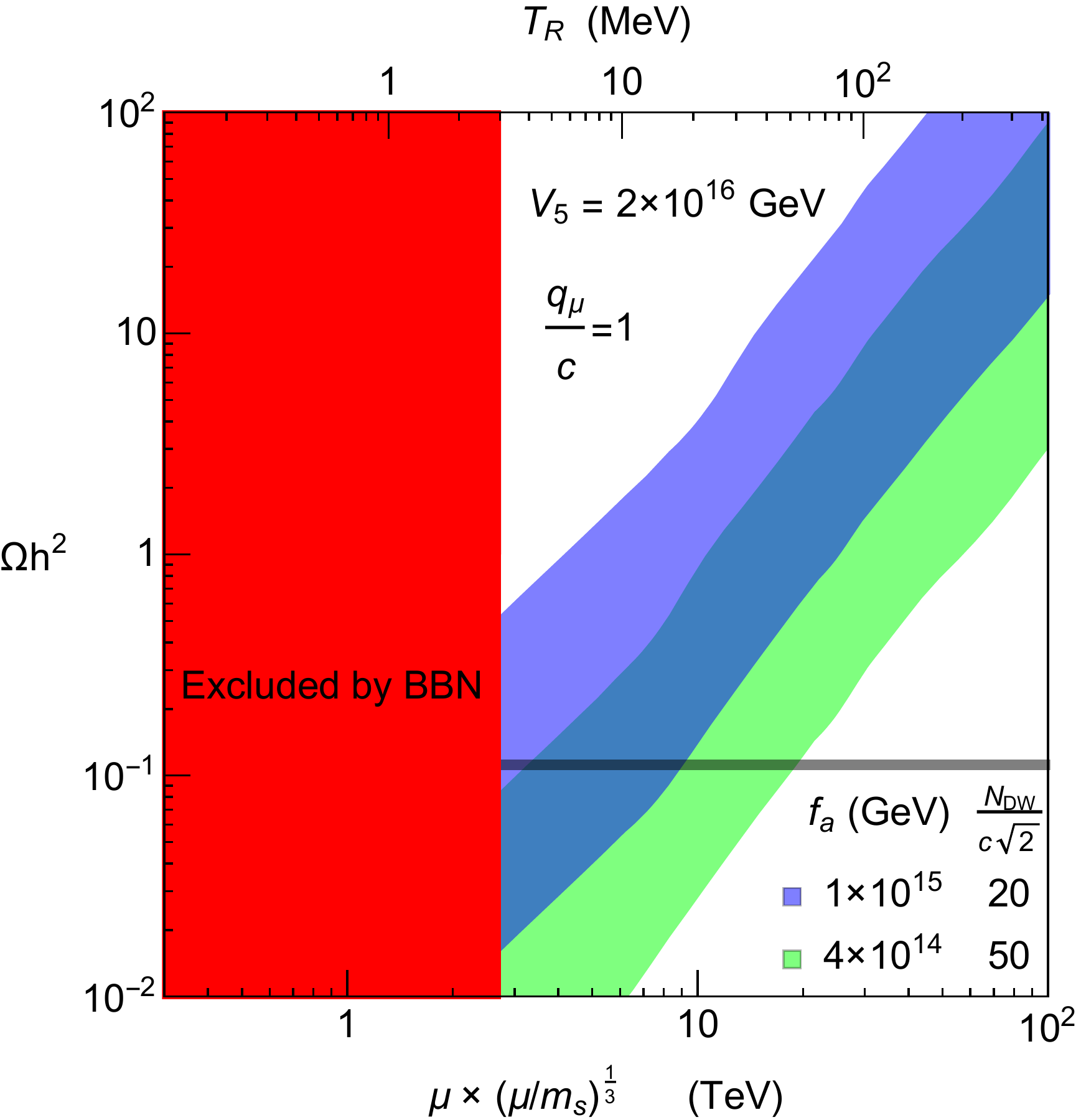}  \includegraphics[width=0.495\textwidth]{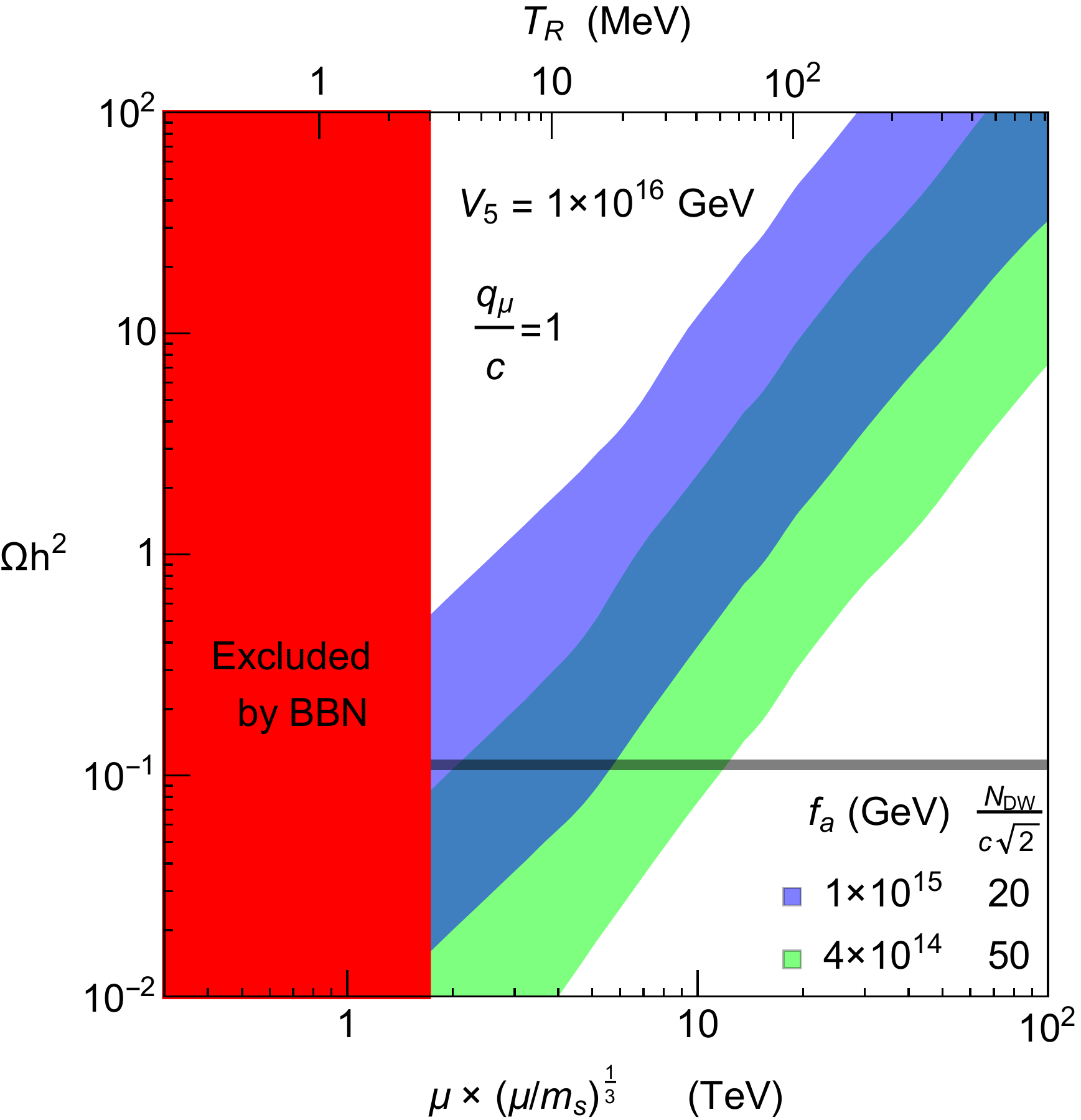} \\ \vspace{0.5cm}
\includegraphics[width=0.495\textwidth]{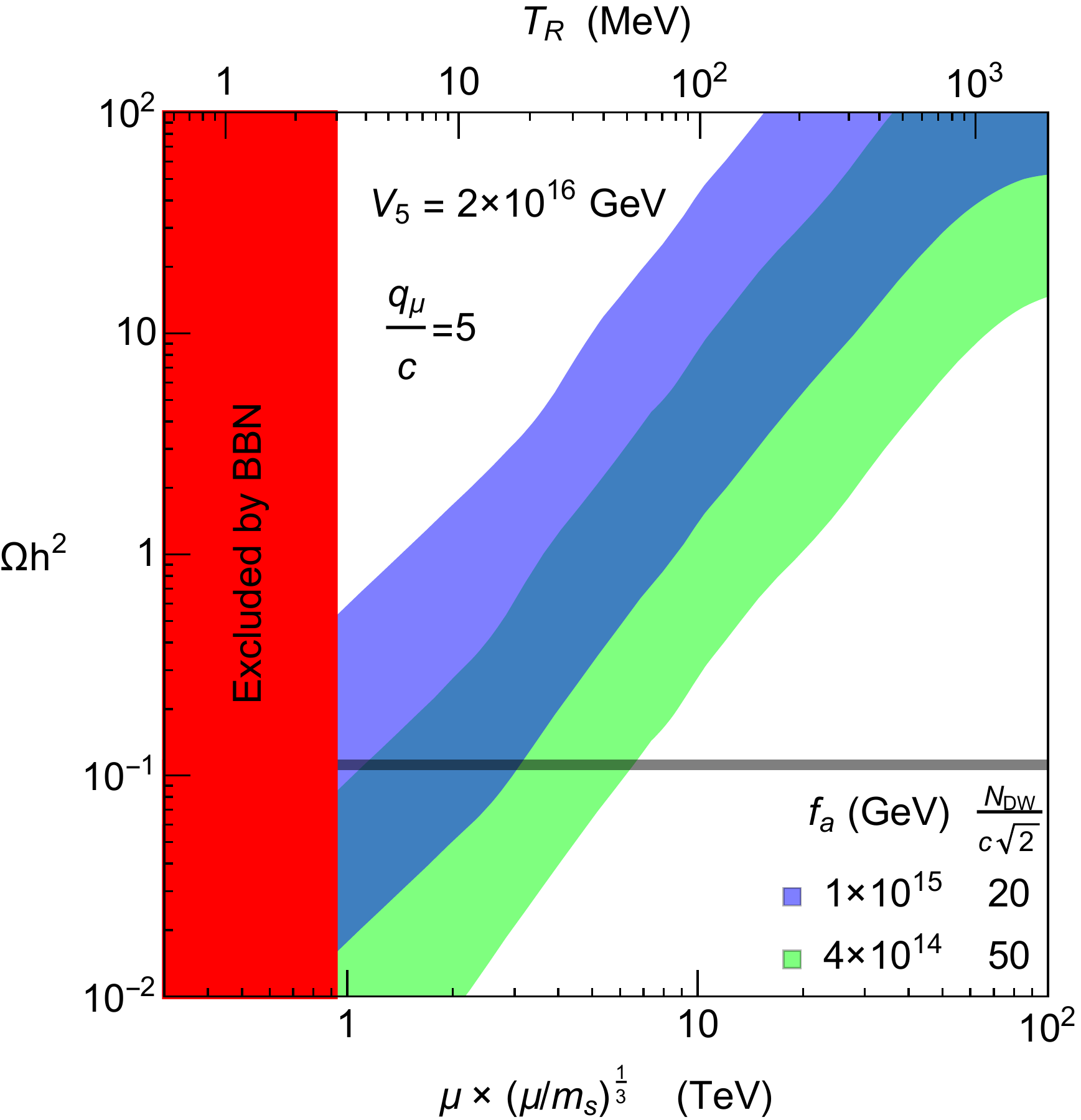} \includegraphics[width=0.495\textwidth]{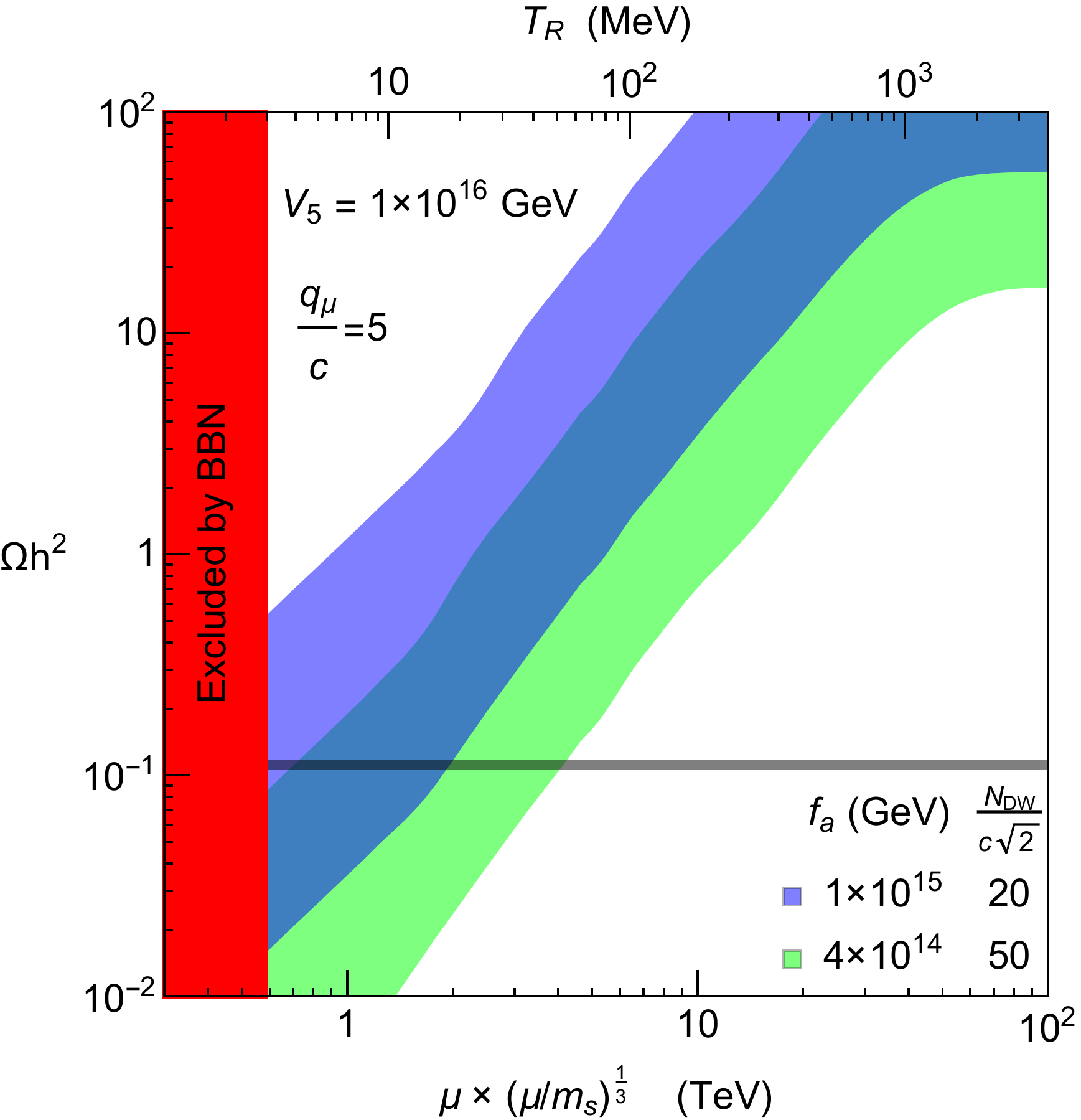}
\end{center}
\caption{Axion dark matter abundance as a function of $\mu(\mu/m_s)^{1/3}$. The horizontal gray band is the observed abundance, while the color bands correspond to the 1$\sigma$ range of $\theta_{i}$ given in Eq.~(\ref{eq:ThetaOneSigma}). Here $\mu/m_s$ has been set to 1 for calculating the phase space of the saxion decay and we assume the saxion can only decay to the SM Higgs and gauge bosons, i.e. $\mathcal{D}=4$.}
\label{fig:OmegaProbability}
\end{figure}

Since PQ breaking occurs before inflation, the initial axion field misalignment angle $\theta_{i}$ has a flat prior distribution between $-\pi$ and $\pi$. As a result, the probability of having $ \left| \theta_{i} \right| \le x \pi$ (for $0 \le x \le 1$) is $x$; in other words, the medium value of $\theta_{i}$ is $\pi/2$. In particular, the range of $\theta_{i}$  
\be
\left( \frac{1}{2} - \frac{1}{2\sqrt{3}} \right) \pi \le \theta_{i} \le   \left( \frac{1}{2} + \frac{1}{2\sqrt{3}} \right) \pi    
\label{eq:ThetaOneSigma}
\ee
corresponds to 1$\sigma$ deviation around the medium value, covering 58\% probability. Using this range, and trading $T_R$ for $\mu(\mu/m_s)^{1/3}$ using Eq.~(\ref{eq:SaxTR}) for specified $q_\mu/c$, we can turn our results into a prediction for $\Omega_a h^2$ as a function of $\mu(\mu/m_s)^{1/3}$ for $V_{PQ} = c V_5$, as shown in Fig.~\ref{fig:OmegaProbability}.  Hence $\mu$ is expected to be of order $(1-10)$ TeV unless $\theta_{i}$ is fine-tuned. The lower panels show that larger values of $q_\mu/c$ allow $\mu$ less than 1 TeV.  It is important to note that this prediction is insensitive to the $\mu/m_s$ ratio, due to its appearance as the 1/3 power.

\section{Dark Radiation}
\label{sec:DarkRad}

\begin{figure}
\begin{center}
\includegraphics[width=0.5\textwidth]{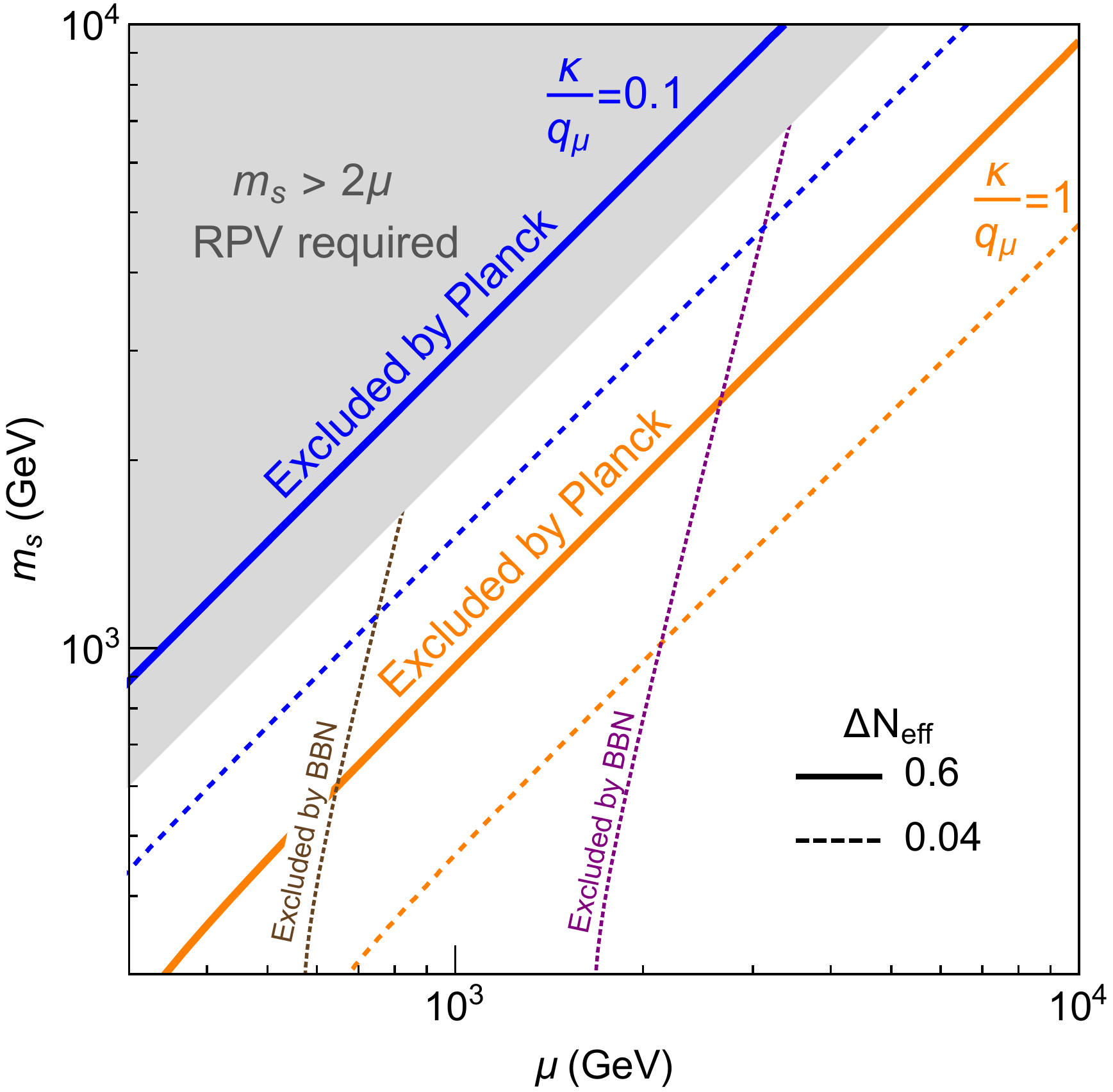} 
\end{center}
\caption{Regions above the bold orange (blue) lines are excluded by the Planck limit of $\Delta N_{eff} < 0.6$, for $\kappa/q_\mu =1 \, (0.1)$.  Future Stage-IV CMB experiments reaching a sensitivity of $ \Delta N_{eff}=0.04$ will probe the theory to the corresponding dashed lines.  The saxion decay branching ratios are computed with $\mathcal{D}=4$. Dotted lines label the BBN bound ($T_R=3$ MeV) for $V_{5}=2\times10^{16}$ GeV and $q_\mu/c=(1,5)$ (purple, brown). In the light gray regions, R parity violation is required to avoid an overabundance of Higgsino dark matter from saxion decay.}
\label{fig:Neff}
\begin{center}
\includegraphics[width=0.55\textwidth]{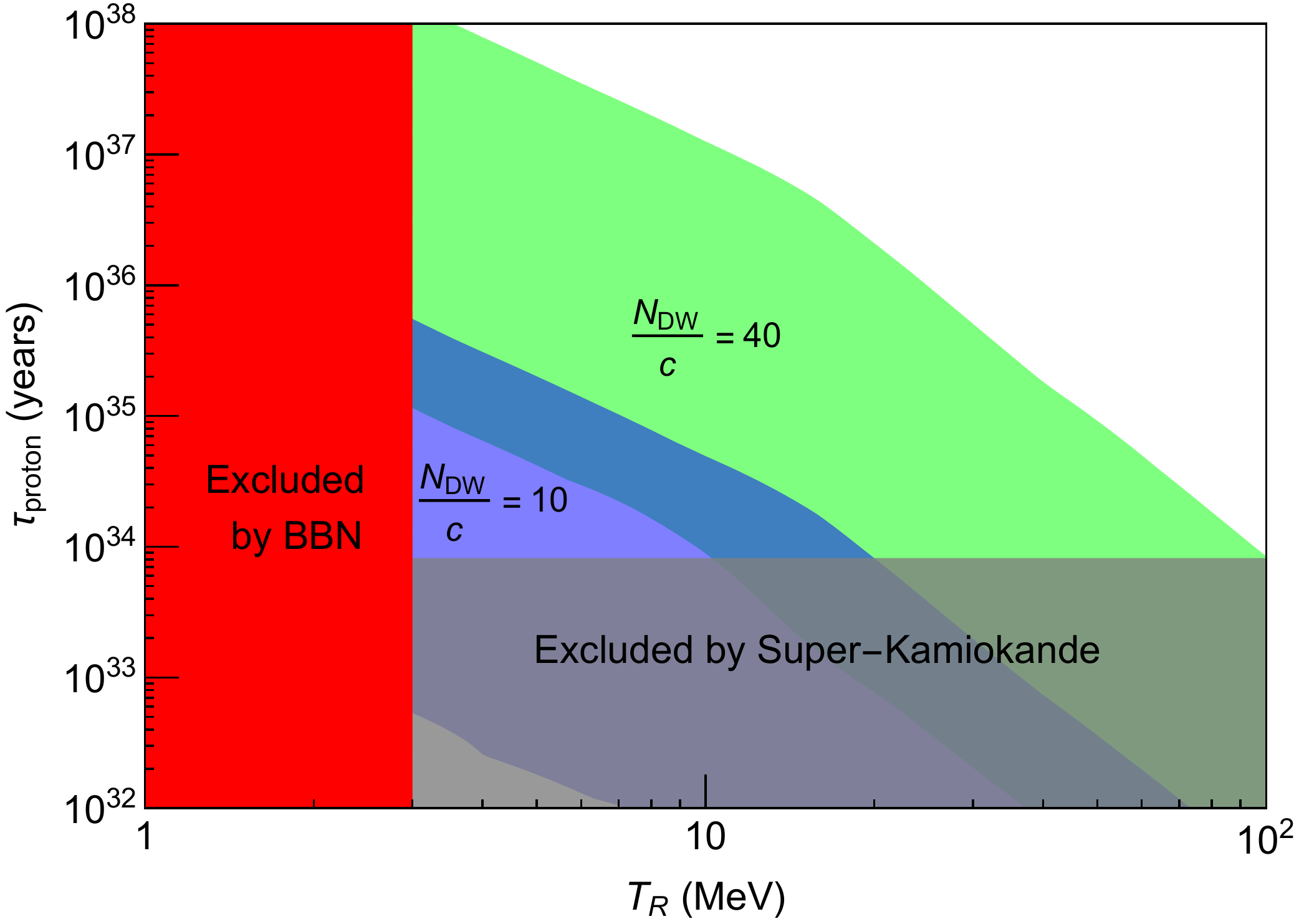}
\end{center}
\caption{The proton lifetime $\tau_p = \Gamma^{-1}(p \rightarrow e^+ \pi^0)$ from $X$ gauge boson exchange as a function of $T_R$ for $N_{\rm DW}/c = 10, 40$. The gray horizontal band is excluded by Super-Kamiokande. Each color band is created using the 1$\sigma$ range of $\theta_{i}$ given in Eq.~(\ref{eq:ThetaOneSigma}).}
\label{fig:ProtonLifetime}
\end{figure}

The decays of the saxion to the axion will contribute to dark radiation \cite{Marsh:2015xka}. The effective number of neutrino species $N_{eff}$ is defined as follows.
\begin{align}
\rho_R = \ & \rho_\gamma + \rho_\nu + \rho_{a} = \left( 1+    \frac{7}{8} \left( \frac{4}{11}  \right)^{4/3}    N_{eff} \right) \rho_\gamma \\
\label{eq:Neff}
\Delta N_{eff} \equiv \  & N_{eff} - N_\nu= \   \frac{8}{7} \left( \frac{11}{4}  \right)^{ \scalebox{1.01}{$\frac{4}{3}$} }   \frac{\rho_{a}}{\rho_\nu}  \frac{\rho_{\nu}}{\rho_\gamma} = \ N_{\nu} \   \frac{\rho_{a}}{\rho_\nu} = \ 3 \ \frac{\rho_{a}}{\rho_\nu} ,
\end{align}
where we know the number of neutrino species $N_{\nu} = 3$ and $\rho_a$ is the additional contribution from other relativistic species -- the axion in this case. \Eq{eq:Neff} is to be evaluated at $T_{\nu\,dec} \sim 1$ MeV, the temperature of neutrino decoupling. The neutrino energy density can be obtained from that of the SM, 
\be
\label{eq:rhoNeutrino}
\rho_\nu(T) = 3 \times 2 \times \frac{7}{8} \frac{\rho_{SM}(T)}{g_*(T)},
\ee
which holds for $T \ge 1$ MeV. We define $\epsilon_a$ as the ratio of the axion energy density $\rho_a$ to that of the standard model $\rho_{SM}$. We assume the saxion energy only goes to those of the axion and SM via saxion decay. Since the saxion decay is the dominant source of both $\rho_a$ and $\rho_{SM}$, $\epsilon_a$ is simply given by the branching ratios of the saxion, 
\be
\label{eq:epsilona}
\epsilon_a \equiv \frac{\rho_{a}(T_R)}{\rho_{SM}(T_R)} = \frac{\Gamma_{s \rightarrow aa}}{\Gamma_{s \rightarrow \text{visible}}} = \frac{1}{4\mathcal{D}} \left( \frac{\kappa}{q_\mu} \right)^2  \left( \frac{m_s}{\mu} \right)^{4}.
\ee
Furthermore, we ignore the mild change of $g_*$ from $T_R$ to $T_{\nu\,dec}$ and this allows us to use \Eq{eq:epsilona} for $T_{\nu\,dec}$. Using the above axion and neutrino energy densities, one finds
\begin{align}
\Delta N_{eff} = & \ \frac{4 }{7} \ \epsilon_a \, g_*(T_{\nu\,dec}) =  \ \frac{g_*(T_{\nu\,dec})}{7 \mathcal{D}} \left( \frac{\kappa}{q_\mu} \right)^2 \,  \left(\frac{m_s}{\mu} \right)^{4} \\
= & \frac{43}{112}  \left( \frac{4}{\mathcal{D}} \right) \left( \frac{g_*(1\text{ MeV})}{10.75}  \right) \left( \frac{\kappa}{q_\mu} \right)^2 \,  \left(\frac{m_s}{\mu} \right)^{4} .
\end{align}
The Planck experimental bound \cite{Ade:2015xua} is $\Delta N_{eff} < 0.6$, which equivalently gives a constraint on the ratio $\mu/m_s$
\be
\frac{\mu}{m_s} > 0.9 \ \left( \frac{\kappa }{q_\mu} \right)^{ \scalebox{1.01}{$\frac{1}{2}$} }  \left( \frac{4}{\mathcal{D}} \right)^{ \scalebox{1.01}{$\frac{1}{4}$} }   \left( \frac{g_*(1 \text{ MeV})}{10.75}  \right)^{ \scalebox{1.01}{$\frac{1}{4}$} } .
\ee
The proposed experiment Stage-IV CMB \cite{Abazajian:2013oma} can be sensitive to $\Delta N_{eff} = 0.04$. This result is plotted in Fig.~\ref{fig:Neff}. 

\section{Proton Decay}
\label{sec:pdecay}

In supersymmetric theories violation of the baryon number can occur via operators of dimensions 4 and 5.  In $SU(5)$ theories superpotential interactions of the form $T \bar{F} \bar{F}$ and $TTT \bar{F}/M_*$ must be very highly suppressed to avoid disastrous proton decay, where $T, \bar{F}$ are the 10 and 5 dimensional representations of matter, even when $M_*$ is taken to be the Planck scale.  In SaxiGUTs the PQ charge of the operator $TTT \bar{F}$ is determined to be $q_\mu$ and since this is necessarily non-zero the PQ symmetry always forbids this operator.  The operator $T \bar{F} \bar{F}$ is also forbidden by the PQ symmetry as long as $\bar{F}$ and $\bar{H}$ have different PQ charges.  Indeed, this PQ charge difference between $\bar{F}$ and $\bar{H}$ provides a distinction between lepton and Higgs doublets, forbidding the operator $H \bar{F}$ and ensuring that $R$ parity is conserved.  Once $PQ$ is spontaneously broken the color triplets in $H$ and $\bar{H}$ acquire a mass and their exchange generates the dimension-5 operator $TTT \bar{F}$.  Even though the corresponding proton decay amplitude is suppressed by small Yukawa couplings, this leads to excessive proton decay in the minimal supersymmetric $SU(5)$ theory.  However, in SaxiGUTs this contribution to proton decay can easily be reduced below the experimental limit by a combination of raising the Higgs triplet mass, raising superpartner masses and flavor suppression.  The resulting signal may be close to the present limit, but is highly model dependent.

Hence we focus on proton decay from dimension-6 operators, induced by the exchange of SU(5) $X$ gauge bosons, that are tightly constrained in SaxiGUTs. The inverse decay rate for $p \rightarrow e^+ \pi^0$ is given by \cite{Babu:2013jba}
\be
\tau_p = \Gamma^{-1}(p \rightarrow e^+ \pi^0) = 1.6\times10^{35} \text{ yrs } \left( \frac{0.012 \text{ GeV}^3}{\alpha_H} \right)^{2} \left( \frac{1/25}{\alpha_G} \right)^{2} \left( \frac{2.5}{A_R} \right)^2 \left( \frac{M_X}{10^{16}\text{ GeV}} \right)^4
\label{eq:TaoProton}
\ee
where $\alpha_H \simeq 0.01 \text{ GeV}^3$ is the nuclear matrix element relevant for proton decay, $\alpha_G = g_5^2/4\pi$ with $g_5$ the unified gauge coupling.  The renormalization factor of the effective dimension-6 proton decay operator is $A_R$, and $M_X$ is the mass of $X$
\be
M_X = \, g_5 V_5 = \, g_5 \frac{V_{PQ}}{c} =  \,  \frac{g_5 N_{\rm DW}}{\sqrt{2} \,c} f_a.
\label{eq:MXvsVGUT}
\ee 
A probabilistic prediction of this decay width is obtained by converting the $f_a$ axis of Fig.~\ref{fig:MisalignContours} into $\tau_p$ using Eqs.~(\ref{eq:TaoProton}) and (\ref{eq:MXvsVGUT}). The result is given in Fig.~\ref{fig:ProtonLifetime} for $g_5 = 0.7$, which also includes the experiment bound of $8.2\times10^{33}$ years from Super-Kamiokande \cite{Nishino:2012bnw}.

\section{Isocurvature Perturbations and CMB Tensor Modes}
\label{sec:isocurvpert}

In SaxiGUTs both PQ and unified gauge symmetries are broken at a scale near $2 \times 10^{16}$ GeV.  Here we investigate whether the energy scale of inflation, $E_I = \rho_I^{1/4}$, could also be of this size so that unified, axion and inflation physics all occur at the same scale.  This is further motivated by the observation that if the observed tilt, $n_s-1$, in the scalar density perturbations, $A_s$, has a significant contribution from the slow roll parameter $\epsilon$, requiring $\epsilon \sim 10^{-2}$, then consistency with the magnitude of $A_s$ leads to $E_I \sim 2 \times10^{16}$ GeV.

The energy scale of inflation is a key quantity for two cosmological observables, the ratio of tensor to scalar density perturbations
\be
r \, = \, \frac{A_t}{A_s} \, \simeq \, 0.2 \left( \frac{E_I}{2\times 10^{16} \GeV} \right)^4
\label{eq:r}
\ee
and the isocurvature density perturbation in theories with axion dark matter and PQ symmetry broken during inflation 
\be
P_{iso} = \left( \frac{ 2 \delta \theta}{\theta_{i}} \right)^2 \hspace{0.25in} \mbox{with} \hspace{0.25in} \delta \theta \simeq \frac{E_I^2}{2 \pi \sqrt{3} f_a^I M_{Pl}}
\label{eq:Piso}
\ee
where $f_a^I$ is the axion decay constant during inflation and $M_{Pl} = 2.4 \times 10^{18}$ GeV.  These two observables arise from quantum fluctuations in the metric and axion fields during inflation, and the observational limits are $r < 0.2$ \cite{Ade:2015lrj} and $P_{iso} < 7 \times 10^{-11}$ \cite{Ade:2015lrj}.


In theories of misalignment axion dark matter with PQ symmetry broken before inflation and $f_a^I = f_a$ Eq. (\ref{eq:Piso}) gives $P_{iso} \propto (E_I^2/\theta_i f_a M_{Pl})^2$.  In the absence of axion dilution from the decay of a condensate, all values of $\theta_i$ and $f_a$ that yield the observed dark matter abundance lead to a bound $E_I \ll 10^{16} $ GeV from the observational limit on $P_{iso}$. Equation~(\ref{eq:r}) then implies there is no prospect of discovering $B$ modes in the microwave background.  Furthermore, the resulting low value of $\epsilon$ may require a high degree of tuning in the inflaton potential.

 In SaxiGUTs two effects allow much larger values of $E_I$ to be consistent with $P_{iso}$.   First, dilution of the axion abundance by saxion decays allows $\theta_{i}$ of order unity for large values of $f_a$ of order $10^{15}$ GeV.   Second, evolution of the saxion field after inflation allows $f_a^I > f_a$ \cite{Linde:1991km} .  Using $f_a^I = (\sqrt{2} /N_{\rm DW})\sigma_i$, where $\sigma_i$ is the initial saxion field value during inflation, the limit from $P_{iso}$ becomes
\be
 E_I \, < \, 10^{16} \GeV \;  \left( \frac{\theta_{i}}{2} \right)^{ \scalebox{1.01}{$\frac{1}{2}$} } \; \left( \frac{10}{N_{\rm DW}} \right)^{ \scalebox{1.01}{$\frac{1}{2}$} } \left( \frac{\sigma_i}{3.2 \times 10^{18} \GeV} \right)^{ \scalebox{1.01}{$\frac{1}{2}$} }
 \label{eq:EI}
 \ee
allowing $E_I \sim (V_{PQ}, V_5)$ for $\sigma_i \sim M_* \sim 10^{18-19}$ GeV.  From the analysis of Sec.~\ref{sec:cosIC} for the initial condition from inflation, case 2 with $\sigma_i \sim M_*$ allows a larger $E_I$ compared to case 1 with $\sigma_i \sim V_{PQ}$. While observation of $r$ is not possible in conventional theories of axion dark matter, it becomes an exciting prospect for SaxiGUTs.
Furthermore,  SaxiGUTs with $E_I \sim (V_{PQ}, V_5)$ and large $\sigma_i$ predict isocurvature perturbations close to the present limit 
\be
P_{iso}\, \simeq \,  7 \times 10^{-11} \left( \frac{E_I}{10^{16} \, \mbox{GeV}} \right)^4 \left( \frac{2}{\theta_{i}} \right)^2 \left( \frac{N_{\rm DW}}{10} \right)^2 \left( \frac{3.2 \times 10^{18} \GeV}{\sigma_i} \right)^2.
\label{eq:Piso2}
\ee

\section{A Summary of Predictions}
\label{sec:con}
We have introduced a class of supersymmetric grand unified theories having PQ symmetry broken at $V_{PQ} = c V_5$, where $V_5$ is the $SU(5)$ breaking scale of gauge coupling unification and $c$ is order unity.  The PQ phase transition occurs before inflation, which sets the saxion field to a value generically of order $V_{PQ}$ or larger relative to its present value.  This leads to an era of saxion oscillations with a reheat temperature of order 10 MeV, implying that axion oscillations from misalignment occur in a matter-dominated era and are significantly diluted.  SaxiGUTs predict the observed dark matter abundance for a supersymmetry breaking scale of order 1-10 TeV and a domain wall number $N_{\rm DW} \sim 10-100$, typical for such unified theories.  These and other predictions are summarized below.

\begin{itemize}
\item {\it A low reheat temperature.}
The decay of the saxion condensate via interactions proportional to the $\mu$ term, which has non-zero PQ charge $q_\mu$, leads to a reheat temperature
\be
T_R \simeq \ 3.5 \ \text{MeV} \ \frac{q_\mu}{c} \left(\frac{\mathcal{D}}{4} \right)^{ \scalebox{1.01}{$\frac{1}{2}$} }  
\left( \frac{\mu}{3 \text{ TeV}} \right)^{ \scalebox{1.01}{$ \frac{3}{2}$} }   \left( \frac{\mu}{m_s} \right)^{\scalebox{1.01}{$\frac{1}{2}$} } \left( \frac{2 \times 10^{16} \ \text{GeV}}{V_{5}} \right) 
\label{eq:TR}
\ee
where $\mathcal{D}$ counts the number of final states in saxion decay and is 4 for decays to $h$, $W^\pm$ and $Z$ bosons, and $m_s$ is the saxion mass.  This correlation between $T_R$ and $\mu$ is shown in the right panel of Fig. \ref{fig:MD_NACondition}.
\item {\it High $f_a$.}
The axion dark matter abundance is closely related to $f_a$, as shown in Eq.~(\ref{eq:Omegah2}), as $f_a$ determines the present value of the axion mass. The resulting prediction for $f_a$ is shown in Fig. \ref{fig:MisalignContours}.  For $\theta_i \simeq 1$, $f_a$ varies in the interval $(0.2-2) \times 10^{15}$ GeV as $T_R$ decreases from 100 MeV to 3 MeV.  However, consistency with $V_{PQ} \sim V_5$ favors low $T_R$, and for $N_{DW}/c <50$ we find the $1\sigma$ range of $\theta_i$ gives $f_a$ in the range $(0.4 - 3) \times 10^{15}$ GeV.  Smaller $\theta_i$ allows larger $f_a$: $\theta_i = 0.1$ allows $f_a$ as large as $10^{16}$ GeV.

Axion dark matter leads to oscillating nuclear electric dipole moments that could be detected via nuclear spin precession \cite{Budker:2013hfa}.  With established techniques, phase 2 of the proposed CASPEr experiment could probe $f_a \gsim 6 \times 10^{15}$ GeV, and the entire range of interest to SaxiGUTs could be tested in further experiments if technical challenges can be overcome \cite{Budker:2013hfa}. A new detection method was proposed in Ref.~\cite{Kahn:2016aff} by exploiting the axion response in a static magnetic field. This new idea probes the axion coupling to photons, and the estimated reach is $f_a \gsim 10^{13}$ GeV. Black hole superradiance already suggests a limit $f_a \lsim 2 \times 10^{17}$ GeV \cite{Arvanitaki:2009fg, Arvanitaki:2010sy}, and with the discovery of gravitational waves from colliding black holes, the prospect of searching for signals at lower $f_a$ is exciting.  For black hole masses of 2$M_\odot$, Advanced LIGO could probe down to $f_a$ of around $2 \times 10^{16}$ GeV \cite{Arvanitaki:2009fg, Arvanitaki:2010sy}, and if black holes of mass $M_\odot/3$ are produced, future gravity wave detectors could probe $f_a$ as low as $3 \times 10^{15}$ GeV. 
\item {\it Axion dark matter.}    Axion misalignment with initial angle $\theta_i$, followed by dilution from saxion decay, leads to a prediction for the amount of axion dark matter
\be
\Omega_a h^2 \, \simeq 0.11 \, \theta_{i}^2 \, q_\mu  c \left( \frac{\mathcal{D}}{4} \right)^{ \scalebox{1.01}{$ \frac{1}{2} $} }  \left(  \frac{35}{N_{\rm DW}} \right)^2 \left(  \frac{\mu}{3 \TeV} \right)^{\scalebox{1.01}{$ \frac{3}{2} $}}  \left( \frac{\mu}{m_s} \right)^{ \scalebox{1.01}{$ \frac{1}{2} $}} \left( \frac{V_{5}}{2 \times 10^{16} \GeV} \right). 
\label{eq:omegaa}
\ee
For $\theta_i, q_\mu$ and $c$ of order unity, this is consistent with providing the entire observed dark matter abundance for $\mu \sim 1-10$ TeV and $N_{\rm DW} \sim 10-100$.  The numerical result for $\Omega_a h^2$ is shown over the parameter space in Fig. \ref{fig:OmegaProbability}. 
\item {\it TeV-scale supersymmetry.}  
Supersymmetry breaking is connected via electroweak symmetry breaking to $\mu$, which is predicted from Eq.~(\ref{eq:omegaa}) and has leading scaling behavior
\be
\mu \, \simeq \, 3 \,\mbox{TeV} \, \frac{1}{\theta_i \sqrt{q_\mu c}} \left( \frac{N_{\rm DW}}{30} \right).
\label{eq:mu}
\ee
The power law dependence of Eq.~(\ref{eq:omegaa}) on $\mu$ provides a more powerful constraint on raising the supersymmetry breaking scale than is provided by logarithmic gauge coupling unification.
A large misalignment angle in a simple SaxiGUT with low $N_{\rm DW}$ may lead to signals at Run-2 of the LHC. As $\mu$ is lowered, consistency of Eq.~(\ref{eq:TR}) with BBN requires a raising $q_\mu/cV_5$.
\item {\it Dark Radiation.}   
Saxion decay to axions leads to a contribution to dark radiation that depends on $\kappa/q_\mu$, where $\kappa$ is defined in Eq.~(\ref{eq:kappadef}).  The constraint from Planck of $\Delta N_{eff} < 0.6$ is shown in Fig. \ref{fig:Neff} together with the parameter space that will be probed by future experiments reaching $\Delta N_{eff} < 0.04$.  The region that can be probed is not large; however, it corresponds to the lowest values of $\mu$ currently allowed (for any fixed values of the other parameters).

\item {\it Proton Decay.}  
The prediction for $\Gamma^{-1}(p \rightarrow e^+ \pi^0)$ from $X$ gauge boson exchange as a function of $T_R$ is shown in Fig. \ref{fig:ProtonLifetime} for the $1 \sigma$ range of $\theta_i$ and $\alpha_G = 1/25$.  This prediction is obtained by normalizing $M_X$ to the observed abundance of dark matter, {\it not} from gauge coupling unification.   The only other parameter relevant for this prediction is $N_{DW}/c$.  For $N_{DW}/c < 10$ and the $1 \sigma$ range of $\theta_i$, consistency between the dark matter abundance and gauge coupling unification requires $T_R \lsim 10$ MeV, and Fig. \ref{fig:ProtonLifetime} shows that such parameters will give a signal in the next generation of experiments searching for proton decay.
\item {\it Cosmic Microwave Background.}  
SaxiGUTs with a large saxion condensate during inflation, $\sigma_i \sim 10^{18-19}$ GeV, allow an energy scale of the vacuum energy during inflation to be of $E_I \sim10^{16}$ GeV, close to the symmetry breaking scales $V_5$ and $V_{PQ}$. Future experiments may then discover both isocurvature density perturbations that arose during inflation from quantum fluctuations in the axion field, as in Eq.~(\ref{eq:r}), and $B$-mode polarization of the CMB radiation, as in Eq.~(\ref{eq:Piso2}). 
\end{itemize}

These diverse phenomena are correlated because SaxiGUTs provides a unified framework for dark matter, the strong CP problem, the TeV scale, gauge unification and inflation.

\section*{Acknowledgments}
We thank Michael Dine, Keisuke Harigaya and Hitoshi Murayama for useful discussions.  This work was supported in part by the Director, Office of Science, Office of High Energy and Nuclear Physics, of the US Department of Energy under Contract DE-AC02-05CH11231 and by the National Science Foundation under grants PHY-1002399 and PHY-1316783. R.C. is supported by the National Science Foundation Graduate Research Fellowship under Grant No. DGE 1106400. F.D. is supported by
the U.S. Department of Energy grant number DE-SC0010107.

\appendix

\section{Axion Supermultiplet Interactions and Saxion Decays}
\label{app:saxion}

In this Appendix we give details about the EFT introduced in Sec.~\ref{sec:EFT}. We give the EFT Lagrangian and then we use these interactions to compute the saxion decay widths given in the main text of this work. The interactions of the axion supermultiplet described by \Eq{eq:aexpansion} of Sec.~\ref{sec:EFT} must respect the shift symmetry described by \Eq{eq:bAshift}. We divide the interactions into three categories: K\"ahler potential, superpotential and SUSY breaking.
 
We work in the basis where the fields $\bPhi_i$ are canonically normalized
\be
\bK = \sum_i \bPhi_i^\dag \bPhi_i  \ .
\label{eq:Kaler}
\ee
The K\"ahler potential interactions for the axion superfield $\bA$ are derived by substituting the PQ breaking fields expansion~(\ref{eq:bPhi}) around the vacuum into \Eq{eq:Kaler}
\be
\bK = \sum_i v_i^2 \exp\left[ q_i \left( \frac{\bA + \bA^\dag}{V_{PQ}} \right) \right] =
 \bA^\dag \bA + \frac{1}{2} \sum_i \frac{q^3_i v_i^2}{V_{PQ}^3} \bA^\dag \bA  \, (\bA + \bA^\dag) + \ldots \ .
\ee
The first equality does not contain any approximation, and we observe that the K\"ahler potential depends only on the combination $\bA + \bA^\dag$, consistently with the shift symmetry in \Eq{eq:bAshift}. The second equality involves the Taylor expansion up to cubic terms. The quadratic piece ensures that the axion superfield is canonically normalized. The cubic terms gives the interaction in \Eq{eq:Lsaa} responsible for the saxion decay to two axions with decay width as in \Eq{eq:Gammasaa}.

Superpotential interactions for the axion superfield alone are forbidden. Holomorphy imposes that we can only have functions of $\bA$ and not $\bA^\dag$, which are not PQ invariant. However, we are interested in DFSZ theories where the combination $\bHu \bHd$ is PQ charged and therefore a $\mu$ is obtained only through PQ breaking. Defining $q_\mu$ as the PQ charge of the $\mu$ term, we have the superpotential interaction
\be
\bW = \mu \exp\left[ q_\mu \frac{\bA}{V_{PQ}} \right] \bHu \bHd =  \mu \bHu \bHd  + q_\mu \frac{\mu}{V_{PQ}} \, \bA \, \bHu \bHd  + \ldots \ .
\ee
The cubic term in the Taylor expansion induces saxion decays to Higgs bosons through the scalar potential interactions
\be
V^{(\rm SUSY)}_{s H_u H_d} = \sqrt{2} \, q_\mu  \frac{\mu^2}{V_{PQ}} \, s \left(H_u^\dag H_u + H_d^\dag H_d  \right) \ .
\label{eq:sHHSUSY}
\ee

Finally, a soft SUSY breaking $B \mu$ term must be present in our theory in order to break the electroweak symmetry. Moreover, this scalar potential term alone also violates PQ. We introduce the SUSY breaking spurion superfield $\bX = \theta^2 B \mu$, and we write down the PQ invariant and SUSY breaking interaction
\be
\bW_{\rm \cancel{SUSY}} = - \bX \, \exp\left[ q_\mu \frac{\bA}{V_{PQ}} \right] \bHu \bHd \ .
\label{eq:WSUSYbreaking}
\ee
Neglecting for a moment the axion superfield, this operator induces the B term
\be
V_{\rm \cancel{SUSY}} = - \int d^2 \theta \bW_{\rm \cancel{SUSY}}  + {\rm h.c.} = B \mu \, H_u H_d + {\rm h.c.}  
\ee
The same operator also generates an interaction between the saxion and two Higgs bosons. As usual, we Taylor expand to linear terms in the $\bA$ superfield and we find the scalar potential contribution
\be
V^{(\rm \cancel{SUSY})}_{s H_u H_d} = q_\mu \,  \frac{B \mu}{V_{PQ}} \, \frac{s}{\sqrt{2}} \left( H_u H_d  + {\rm h.c.}  \right) \ .
\label{eq:sHHSUSYbreaking}
\ee

The visible decay channels for the saxion are to Higgs bosons. The Higgs doublets $H_u$ and $H_d$ contain a total of $8$ real scalar degrees of freedom, with $3$ Goldstone bosons ($G^\pm$ and $G^0$) eaten up by the EW gauge bosons. The remaining spectrum consists of two CP-even neutral scalar ($h$ and $H$), one CP-odd neutral scalar ($A$) and one charged scalar ($H^\pm$). We organize these degrees of freedom by introducing the two doublets
\begin{align}
H_{\rm SM} = & \, \left( \begin{array}{c} G^+ \\ v + \frac{h + i G^0}{\sqrt{2}}  \end{array} \right) \ , \\
H_{\rm Heavy} = & \, \left( \begin{array}{l} \frac{H + i A}{\sqrt{2}}   \\   \, H^- \end{array} \right) \ .
\end{align}
The field $H_{\rm SM}$ is just the SM Higgs doublet, which includes the EWSB vev $v = 174 \, {\rm GeV}$. The extra scalars all reside inside $H_{\rm Heavy}$. The pseudoscalar mass 
\be
m_A^2 = \frac{2 B \mu}{\sin2\beta} \ ,
\label{eq:mA}
\ee
with $\tan\beta = v_u / v_d$, is the parameter controlling how far we are from recovering the SM. In the decoupling limit $m_A \gg m_Z$, which is always the case we are interested in, the gauge eigenstates can be compactly expressed in terms of the mass eigenstates
\begin{align}
H_u =  & \, \sin\beta \, H_{\rm SM} \, + \cos\beta \, \widetilde{H}_{\rm Heavy} \ , \\
H_d = & \, \cos\beta \, \widetilde{H}_{\rm SM} \, +  \sin\beta \, H_{\rm Heavy}\ ,
\end{align}
where we define $ \widetilde{H}_{\rm SM} = i \sigma^2 H_{\rm SM}^*$ and the same for $\widetilde{H}_{\rm Heavy}$. 

Saxion visible decays are described by the operators in \Eqs{eq:sHHSUSY}{eq:sHHSUSYbreaking}. We express these interactions in terms of the mass eigenstates, and the final results of this procedure read
\be
\begin{split}
V_{s H H} = & \, \sqrt{2} \, q_\mu  \frac{\mu^2}{v} \, s 
\left[ H_{\rm SM}^\dag H_{\rm SM} + H_{\rm heavy}^\dag H_{\rm heavy} \right] +  \\ &
- q_\mu \,  \frac{B \mu}{v} \, \frac{s}{\sqrt{2}} \sin2\beta 
\left[H_{\rm SM}^\dag H_{\rm SM} - H_{\rm heavy}^\dag H_{\rm heavy} +
\frac{1}{\tan2\beta} \left( H_{\rm SM} H_{\rm heavy} + {\rm h.c.} \right)\right] \ .
\end{split}
\label{eq:VsHHFINAL}
\ee
The first and second rows give the SUSY preserving and breaking contributions, respectively. There are three different types of decays to Higgs bosons, according to the multiplicity of SM particles in the final state, which can range from zero to two. For decays to SM final states 
\be
\label{eq:GammaSaxhh}
\Gamma_{s \, \rightarrow \, h h} =  \Gamma_{s \, \rightarrow \, Z Z} = \frac{1}{2} \Gamma_{s \, \rightarrow \, W W}  = \frac{q^2_\mu \mu^4}{16 \pi m_s V_{PQ}^2} \left( 1 - \frac{1}{4} \frac{m_A^2}{\mu^2} \sin^22\beta \right)^2   \ ,
\ee
where we have used \Eq{eq:mA} to trade $B$ with $m_A$. Here, decays to electroweak gauge bosons are evaluated in the Goldstone equivalence limit. Likewise, the decay to a pair of heavy Higgs bosons can be computed from the interactions in \Eq{eq:VsHHFINAL}, giving a decay width
\be
\Gamma_{s \, \rightarrow \, H H} = \Gamma_{s \, \rightarrow \, A A} = \frac{1}{2} \Gamma_{s \, \rightarrow \, H^+ H^-} = \frac{q^2_\mu \mu^4}{16 \pi m_s V_{PQ}^2} \left( 1 + \frac{1}{4} \frac{m_A^2}{\mu^2} \sin^22\beta \right)^2   \ .
\label{eq:GammaSaxHH}
\ee
Finally, the saxion has also the option of decaying to final states composed of one SM particle and one heavy Higgs boson. The decay widths for these cases read
\be
\label{eq:GammaSaxhH}
\Gamma_{s \, \rightarrow \, h H}  = \Gamma_{s \, \rightarrow \, A Z} = \Gamma_{s \, \rightarrow \, W^+ H^-} =
\Gamma_{s \, \rightarrow \, W^- H^+} =  \frac{q^2_\mu \, m_A^4}{512 \pi m_s V_{PQ}^2} \sin^24\beta   \ .
\ee

To summarize, saxion decay rates to visible matter are quantified by Eqs.~(\ref{eq:GammaSaxhh}), (\ref{eq:GammaSaxHH}) and (\ref{eq:GammaSaxhH}). These expressions are completely general and they only assume the validity of the decoupling limit regime $m_A \gg m_Z$. The total width for visible channels depends on the saxion mass with respect to $m_A$, which determines how many channels are kinematically available. For saxion mass values such that only SM final states are possible we have the saxion visible width
\be
\left. \Gamma_{s \, \rightarrow \, {\rm visible}} \right|_{\mathcal{D} = 4} =  \frac{q^2_\mu \mu^4}{4 \pi m_s V_{PQ}^2} \left( 1 - \frac{1}{4} \frac{m_A^2}{\mu^2} \sin^22\beta \right)^2  \ , \qquad \qquad \qquad m_s \lesssim 2 m_A \ .
\ee 
Here, $\mathcal{D} = 4$ denotes the number of allowed particles in the final states. On the contrary, for saxion mass above the heavy Higgs bosons threshold we have the saxion visible width
\be
\left. \Gamma_{s \, \rightarrow \, {\rm visible}} \right|_{\mathcal{D} = 8} = \frac{q^2_\mu \mu^4}{2 \pi m_s V_{PQ}^2} 
\left( 1 + \frac{m_A^4}{16 \, \mu^4}   \sin^22\beta \right) \ , \qquad \qquad \qquad m_s \gtrsim 2 m_A \ .
\ee
In this work we use the large $\tan\beta$ limit of these expressions, which gives \Eq{eq:GammaSaxionVisible}.

\end{document}